\def\puncspace{\ifmmode\,\else{\ifcat.\C{\if.\C\else\if,\C\else\if?\C\else%
\if:\C\else\if;\C\else\if-\C\else\if)\C\else\if/\C\else\if]\C\else\if'\C%
\else\space\fi\fi\fi\fi\fi\fi\fi\fi\fi\fi}%
\else\if\empty\C\else\if\space\C\else\space\fi\fi\fi}\fi}  
\def\SP{\let\\=\empty\futurelet\C\puncspace}  
\def\etal{et al.}

\def\h{${{\rm h}_{75}}^{-1}$}  
\def\lsim{~\rlap{$<$}{\lower 1.0ex\hbox{$\sim$}}}  
\def\gsim{~\rlap{$>$}{\lower 1.0ex\hbox{$\sim$}}}  
\def\void#1{{}}

\documentclass{aa}  
\usepackage{graphicx}

\def\etal{{\it et al.\/}\ }  
  
\begin{document}  
\title {Multiple merging events in Abell 521
\thanks{Based on observations made at the Canada France Hawaii Telescope and at the European Southern Observatory. CFHT is operated by the National Research Council of Canada, the Centre National de la Recherche Scientifique of France, and the University of Hawaii.}}
\author{Ferrari, C.\inst{1}  
\and Maurogordato, S.\inst{1}   
\and Cappi, A.\inst{2,1}  
\and Benoist, C.\inst{1}}  
\offprints{Chiara Ferrari}  
\institute{    
  CERGA, UMR 6527, CNRS, Observatoire de la C\^ote d'Azur, BP4229, Le Mont-Gros, 06304 Nice Cedex 4, France \and INAF, Osservatorio Astronomico di Bologna, via Ranzani 1, 40127 Bologna, Italy   
  }  
\date{Received~12 August 2002 / Accepted~12 November 2002}  

\abstract{We present a detailed spatial and dynamical analysis of 
the central 
$\sim$~2.2~\h~Mpc region of the galaxy cluster Abell~521 (z=0.247), based on 
238 spectra (of which 191 new measurements) obtained  
at the 3.6~m Telescope of the European Southern Observatory and at 
the Canada-France-Hawaii Telescope.
From the analysis of the 125 galaxies that are confirmed members of the
cluster, we derive a location (``mean'' velocity) of
${\rm C}_{\rm BI}=74019 ^{+112} _{-125}$~km/s and detect a complex
velocity distribution with high velocity scale (``dispersion'', 
${\rm S}_{\rm BI}=1325 ^{+145} _{-100}$ km/s), but clear departure
from a single Gaussian component. When excluding a possible background
group of four galaxies, the velocity dispersion
remains still large ($\sim 1200$ km/s).
The general structure of the
cluster follows a North-West/South-East direction, crossed by a
perpendicular high
density ``ridge'' of galaxies in the core region. 
The Northern region of the cluster is characterized by a lower 
velocity dispersion as compared to the whole cluster value; it hosts
the BCG and a dynamically bound complex of galaxies, and it is 
associated with a group detected in X-ray (Arnaud \etal 2000). 
This region could be in a stage of pre-merger onto the
main cluster. The small offset  (${\sim}+250$ km/s) in the mean
velocity of the northern region as compared to the whole cluster
suggests that the merging occurs partly in the plane of the sky.
These results, taken together with the fact that most of the 
clumps detected on the isodensity maps, as well as 
the early-type galaxies and the brightest ones 
(L$>{\rm L}^*$) are aligned, suggest that this North-West/South-East direction
is the preferred one  for the  formation of this cluster. 
The~central high dense region (``ridge'') shows a lower velocity location 
(${\rm C}_{\rm BI}~=~73625 ^{+344} _{-350}$~km/s) and significantly 
higher scale  ($1780 ^{+234} _{-142}$~km/s) as compared to the whole
cluster values. This is due to the presence of a low-velocity 
group of galaxies with a high fraction of emission line objects. This
can be explained in a scenario in which a merging of subclusters  has
recently occurred along the direction of the ``ridge'' with a
significant component along the line of sight. The low-velocity group
would then be a high-speed remnant of the collision 
which would have also triggered an episode of intense star formation
responsible for the large fraction of late-type objects in this
region.
}

\maketitle   
\section{Introduction}

In the hierarchical model of structure formation, galaxy  clusters are
supposed to form by merging of units of smaller mass. 
Analysis of statistical samples of  galaxy clusters have shown that a high
percentage of clusters with substructures is detected even at low redshift,
implying that clusters are still today undergoing the process of
formation (Geller \& Beers 1982, Dressler \& Shectman 1988, Jones
\& Forman 1992). Moreover, 
quantifying precisely the amount of morphologically complex clusters 
allows one in principle to constrain directly the
cosmological model through the density parameter $\Omega_m$ (Richstone et
al. 1992, Mohr~\etal 1995). This analysis is however hampered by the
existing uncertainty in the rate at which substructure is erased
(Kauffmann \& White 1993, Lacey \& Cole 1993). The second difficulty
is that  sub-clustering
can affect the various quantities observable, such as the
projected distribution of the galaxies and of the
gas, the velocity distribution of the galaxies and the
temperature structure of the gas, not necessarily at the same level,
leading to sometimes different conclusions. 

Detailed studies of individual
complex galaxy clusters at different wavelengths is a complementary
analysis which allows one to obtain details of the scenario of formation
of these objects, and the physical processes necessary to explain
the observed distribution (Flores~\etal 2000, Donnelly~\etal 2001,
Mohr~\etal 1996, Bardelli~\etal 1998, Rose~\etal 2002, Berrington
\etal2002, Czoske~\etal 2002, Valtchanov~\etal 2002). 
In this paper, we will concentrate on the
dynamical analysis of the  merging cluster Abell 521 which has been
targeted for its outstanding properties. 

Abell 521 is a rich (R=1) Abell cluster, first detected in X-ray with
HEAO1 (Johnson~\etal 1983, Kowalski~\etal 1984). It was also suspected
(Ulmer~\etal 1985) to form a binary cluster together with its nearest
neighbor on the sky, A518, but no clear evidence for gas interaction
was found between the two clusters. Radio observations in the region
of this cluster (Hanisch~\etal 1985) have also shown a high fraction of
radio sources with projected distances to the center compatible with
these objects being cluster members. More recent data, both in X-ray
and optical (Arnaud~\etal 2000, Maurogordato~\etal 2000) provided a
more detailed analysis of the properties of the galaxies and gas
distributions in this cluster. Imaging in X-ray (ROSAT/HRI) has
shown a gas morphology with two peaks which can respectively be
associated with a diffuse main cluster, and a compact less massive
group in the northern region, suspected to be in pre-merger stage with
the main cluster. The projected galaxy density distribution in the
central 2.2~\h~Mpc has a very anisotropic morphology, as it
exhibits two high density filaments crossing in an X-shape structure
at the barycentre of the cluster. A severe gas/galaxy segregation 
stands out. The brightest cluster galaxy is offset of the cluster
barycentre, and lies in the region of the X-ray northern
group. Multi-object spectroscopy at ESO/EFOSC2 and CFHT/MOS led to the
determination of the mean redshift of the cluster, z=0.247, and of
its velocity dispersion ${\rm S}_{\rm BI}=1386 $ km/s as measured from 41
members. However, this very high value of the velocity dispersion could
be affected by the presence of substructures. Its value
is also high compared to the temperature of the X-ray gas measured
with ASCA (T=6.3 KeV, Arnaud~\etal 2000). These results imply
that this cluster is undergoing strong dynamical evolution.
This motivated new observations in order to better characterize the
merging scenario, in particular through additional multi-object
spectroscopy. A project of wide-field multicolor imaging in five bands is under progress, and will address the large scale environment of the cluster 
using photometric redshifts (Ferrari~\etal in prep.).
In this paper, we analyze in detail the velocity distribution within
the central 2.2~\h~Mpc of the cluster with new data
obtained at the ESO 3.6~m telescope (191 new redshifts
measured). Section 2 briefly describes the observations and the data
processing techniques, as well as the level of completeness
achieved. In Sect. 3, we perform a general analysis of the 1D velocity
distribution, test for departures from an unimodal gaussian, and fit a
partition in three velocity groups.  In Sect. 4, correlations
between the structures identified in velocity space and in projected
coordinates have been looked for. Variations of dynamical properties 
with absolute luminosity, color, and spectral type have been 
addressed in Sect. 5.
In Sect. 6, our new data are used to elaborate the more plausible
scenario of occurrence of the various merging events within 
this particularly  complex cluster. 
All numbers are expressed as a function of ${\rm h}_{75}$, the Hubble constant
in units of 75~km/s/Mpc. We have used the $\Lambda$CDM model with
$\Omega_{\rm m}=0.3$ and $\Omega_{\Lambda}=0.7$, then
1~arcmin corresponds to $\sim$0.217~\h~Mpc in the following.

\section {The data}  
\subsection {Observations and data reduction}  

\begin{figure*}  
\resizebox{18cm}{!}  
{\includegraphics{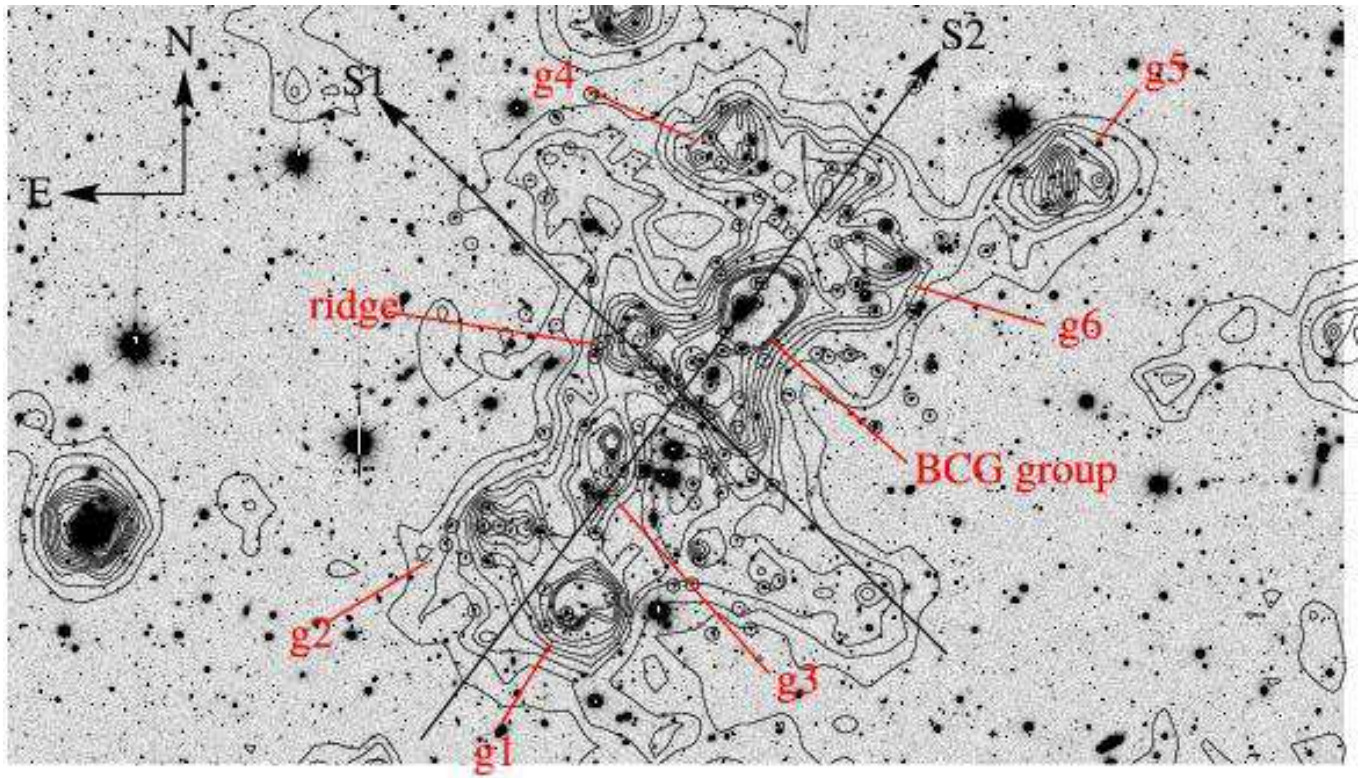}}  
\hfill  
\parbox[b]{18cm}{  
\caption{Central field (20'$\times$10') of the cluster. In black, galaxy iso-density contours for B$<$27 and I$<20$; the lowest iso-density contour corresponds to 1$\sigma$ level above the mean density in the field, the contours are spaced by 0.5$\sigma$. Black circles correspond to the 125 galaxies of the cluster with spectral quality flag=1. The subgroups identified in the projected density map and the two main directions of the cluster (S1 and S2) have been shown.}  
\label{FC}}  
\end{figure*}  
   
New data have been obtained through a campaign of multi-object spectroscopy  
at the ESO 3.6 m telescope (3 nights in October 1999,   
and 2.5 nights in December 2000).   
We used the ESO Faint Object Spectrograph and Camera (EFOSC2)  
with grism\#04, whose grating of 360~line~${\rm mm}^{-1}$ leads 
to a dispersion of   
1.68~\AA/pixel, and a wavelength coverage ranging  from   
4085 to 7520~\AA. The detector used was the EFOSC2 CCD Loral/Lesser\#40, 
with an image size of 2048 $\times$ 2048  
(we made a 2 $\times$ 2 binning, in order to improve the signal to noise   
ratio), and a pixel size of 15 $\times$ 15 $\mu$m.  

During the run of October 1999 we achieved a spectral resolution of 
FWHM$\sim$18.5~\AA, 
while in  
the second run, as a smaller punching head was available    
({$1.35^{\prime\prime}$ instead of {$1.8^{\prime\prime}$),  
the resolution was improved to FWHM$\sim$12.5~\AA.  
The total integrated exposure time was 9000~s for each frame, 
split in at least two exposures to eliminate cosmic rays. After each
science exposure, a Helium-Argon lamp exposure was systematically taken for
wavelength calibration. 

Data have been reduced with IRAF\footnote{IRAF is distributed by the   
National Optical Astronomy Observatories, which are operated by the   
Association of Universities for Research in Astronomy, Inc., under cooperative  agreement with the National Science Foundation}, using our automated package
for multi-object spectroscopy based on the task ``apall''. 
Radial velocities were determined using the cross-correlation  
technique (Tonry \& Davis, 1981) implemented in the RVSAO package   
(developed at the Smithsonian Astrophysical Observatory)   
with radial velocities standards obtained from the observations of   
late-type stars.  
We have obtained 191 new spectra. Among these,  
29 are stars, while 109 are identified as galaxy spectra with a 
signal to noise ratio   
sufficient to obtain radial velocity measurement with a parameter R of   
Tonry \& Davis greater than 3. The remaining 53 objects have
a poor velocity determination.  

We list in Table~\ref{FindC} our new velocity measurements. The   
columns read as follows:   
Col.~1: identification number of each target galaxy; 
Col.~2: run of observations (October 1999=ESO1, December 2000=ESO2); 
Cols.~3 and 4: right ascension and declination (J2000.0) of the target   
galaxy;  
Cols.~5 and 6: best estimate of the radial velocity and associated 
error from the cross-correlation technique (those values have
been set to ``-2'' if  
the object is a star and to ``-1'' if we have no redshift information);  
Col.~7: a quality flag for the redshift determination:   
1=good determination (R~$\geq$~3),  
2=uncertain determination,  
3=very poor determination,  
4=failed spectra,  
Col.~8: a listing of detected emission lines.

\begin{figure}  
\resizebox{8cm}{!}  
{\includegraphics{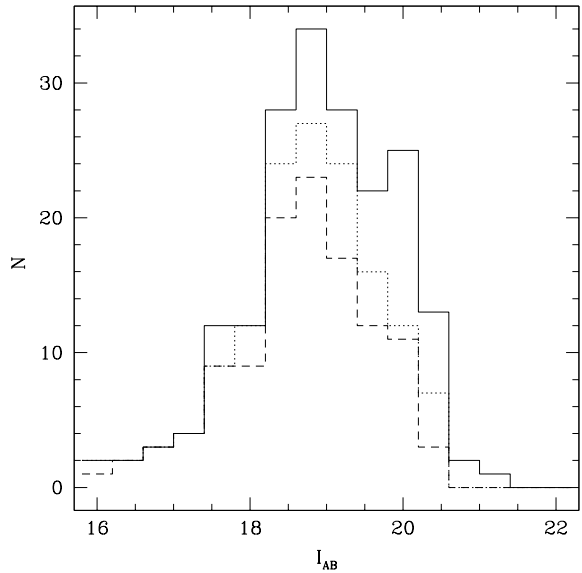}}  
\hfill  
\parbox[b]{8cm}{  
\caption{I-band magnitude distributions of the galaxies of our spectroscopic sample (187 objects - solid line), of all the galaxies with good velocity determination (141 - dot), and, among them, of those belonging to A521 (113 - dash).}  
\label{I}}  
\end{figure}

\begin{figure}  
\resizebox{8cm}{!}  
{\includegraphics{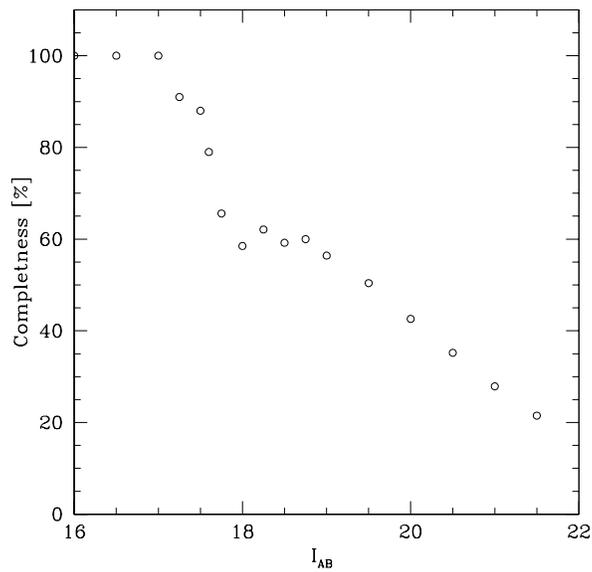}}  
\hfill  
\parbox[b]{8cm}{  
\caption{Velocity completeness for different cuts in I-band magnitude in the central $10{\times}10~{\rm arcmin}^2$ field  
covered by spectroscopy. }  
\label{completI}}  
\end{figure}

\begin{figure} 

\centering

\resizebox{7.5cm}{!}{\includegraphics{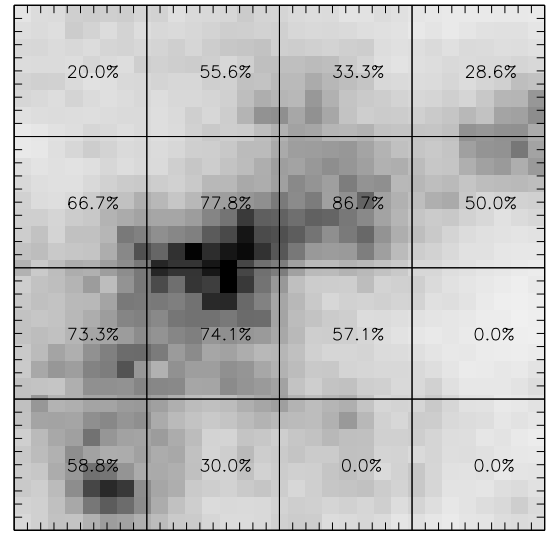}}
 
\resizebox{7.5cm}{!}{\includegraphics{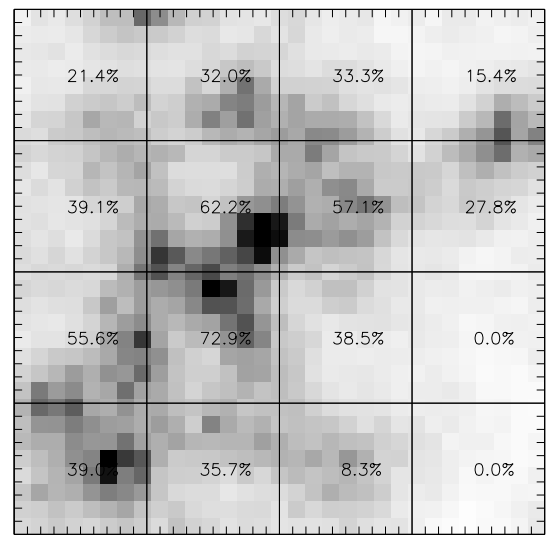}}
  
\resizebox{7.5cm}{!} {\includegraphics{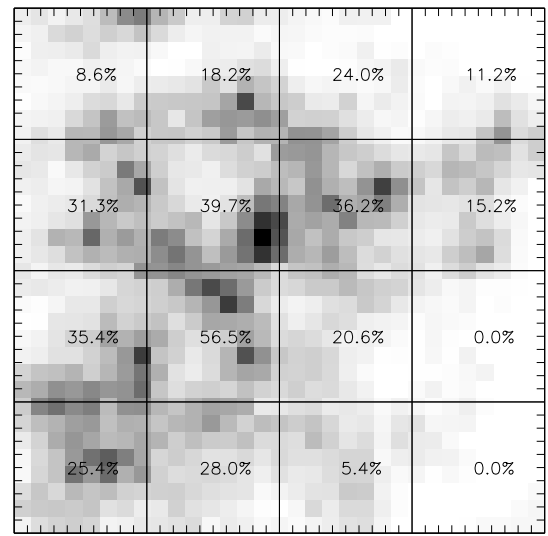}}

\parbox[b]{8cm}{  
\caption{Completeness factor is superimposed to the isodensity maps in the field selected for spectroscopy (10'$\times$10'). Each cell of the grid covers 2.5'$\times$2.5'. Different magnitude cuts have been considered; from top to bottom: I$<$19, I$<$20, I$<$21. }  
\label{griglia}}  
\end{figure}

\subsection{The spectroscopic sample}\label{complet}  
  
In the following analysis, the new set of spectroscopic 
data presented above has
been combined to our original sample (Maurogordato~\etal  2000, 47
objects added to our new catalogue) resulting in a sample of 
209 galaxies and 29 stars. 

B- and I-band imaging (CFH12k) taken as part of our multicolor imaging program
were used to build a color catalogue in the central region of the cluster 
surveyed by spectroscopy. 
This allowed to associate B- and I-band magnitudes for most galaxies
of the spectroscopic sample (187), except for 22 objects located in the gaps between the chips of the camera or 
strongly blended. Fig.~\ref{FC} displays the central field of the cluster, 
where galaxies with secure redshift determination have been circled. 
The isodensity map of the projected distribution of galaxies 
with 
I-band magnitude I$<$20 is also displayed 
(derived using the Dressler algorithm; Dressler 1980). 
The preferential directions S1 and S2 observed in Arnaud~\etal 2000
are indicated. The density structures detected at more than $5 \sigma$ level 
are also indicated: 
six groups called g1 to g6, a group around the BCG, and the so-called 
``ridge'' structure corresponding to S1. 

In Fig.~\ref{I} we have plotted the
I-band magnitude distributions of: the galaxies of our
magnitude/velocity sample (187), those with a very good
redshift determination (141), and, among them, those belonging to
the cluster (113).
In Fig.~\ref{completI} we show the ratio of the number of objects 
with measured velocities to the total number of galaxies detected  within the
central $10{\times}10~{\rm arcmin}^2$ of the field as a function of
the I-band magnitude.   We reach a
general level of completeness for spectroscopy of 50\% at 
${\rm I}_{\rm AB}=19.5$, which drops at
30\% at ${\rm I}_{\rm AB}=20.5$. However, these values are strongly affected by
several incomplete fields at the periphery of the cluster, as
the central dense regions of the cluster have been much better
sampled. In fact, we have divided the spectroscopic
field in $2.5{\times}2.5~{\rm arcmin}^2$ cells (Fig.~\ref{griglia}),
and measured the degree of completeness in each cell for three
different cuts in I magnitude (${\rm I}_{\rm AB}=19,20,21$). The
obtained values are shown in each cell of the
corresponding isodensity maps. At ${\rm I}_{\rm AB}=19$, we have an excellent velocity sampling
(${\sim}75{\div}80$\% completeness) in the North-West/South-East main
structure of the cluster, which drops at ${\sim}60$\% at ${\rm I}_{\rm AB}=20$,
and at ${\sim}40$\% at ${\rm I}_{\rm AB}=21$.

\section{The velocity distribution}  
  
\subsection{Global analysis of the velocity distribution}  
  
We have analyzed the general behavior of the velocity distribution with  
the ROSTAT package (Beers~\etal 1990). For this purpose, and in all the 
following analysis, we have used only the 125 objects with quality   
flag=1 (secure redshift) in our dataset; their velocity histogram
is shown in Fig.~\ref{histoTOT}.

  
In the case of large number of redshifts ($100{\leq}{\rm n}{\leq}200$) 
as in our situation, the best choice   
to estimate location (``mean'' velocity) and scale (velocity ``dispersion'')
is the biweight estimator (Beers~\etal 1990),
as it provides the best combination of resistance and efficiency across the
possible contaminations of a simple gaussian distribution. 
We find a location ${\rm C}_{\rm BI}=74019 ^{+112} _{-125}$~km/s   
and a scale ${\rm S}_{\rm BI}=1325 ^{+145} _{-100}$~km/s. 
These results are in good agreement with those obtained  
from the analysis of our former sample of 41 galaxies (Maurogordato   
\etal 2000), and confirm the apparently high value of the velocity 
dispersion in this cluster.  
   
\begin{figure}  
\resizebox{8cm}{!}{\includegraphics{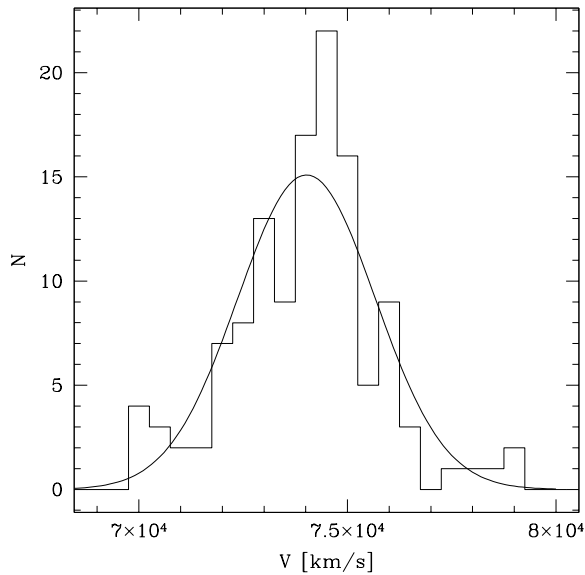}}  
\hfill  
\parbox[b]{8cm}{  
\caption{Velocity histogram within Abell 521 obtained from the 125 
Q.F.=1 members of the cluster, with a binning of 500~km/s. A 
gaussian  
function with the velocity distribution derived with ROSTAT is 
superimposed.}  
\label{histoTOT}}  
\end{figure}  

\begin{figure}  
\resizebox{8cm}{!}{\includegraphics{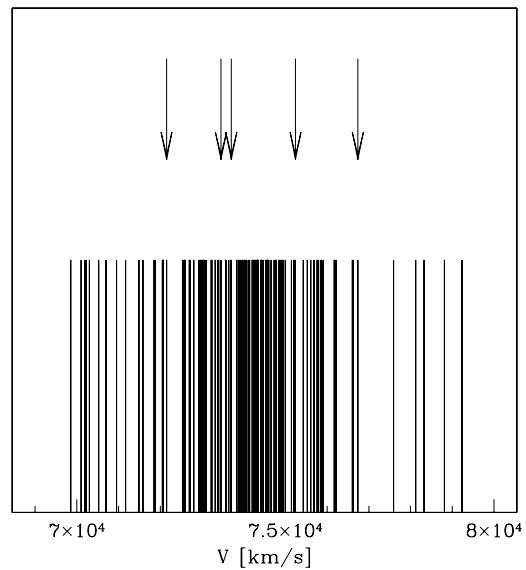}}  
\hfill  
\parbox[b]{8cm}{  
\caption{Stripe density plot of radial velocities for the 125 members of Abell 521.}  
\label{stripe}}  
\end{figure}  

\begin{figure}  
\resizebox{8cm}{!}{\includegraphics{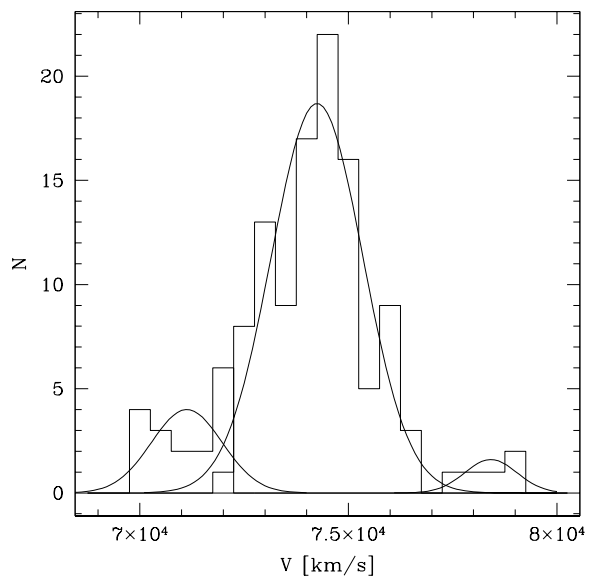}}  
\hfill  
\parbox[b]{8cm}{  
\caption{Same as Fig.~\ref{histoTOT}, but now the best fit Gaussians found by  KMM for a three group partition is superimposed and we have used a binning of 500 km/s.}  
\label{KMM}}  
\end{figure}

\begin{figure}
\centering  
\resizebox{7.3cm}{!}{\includegraphics{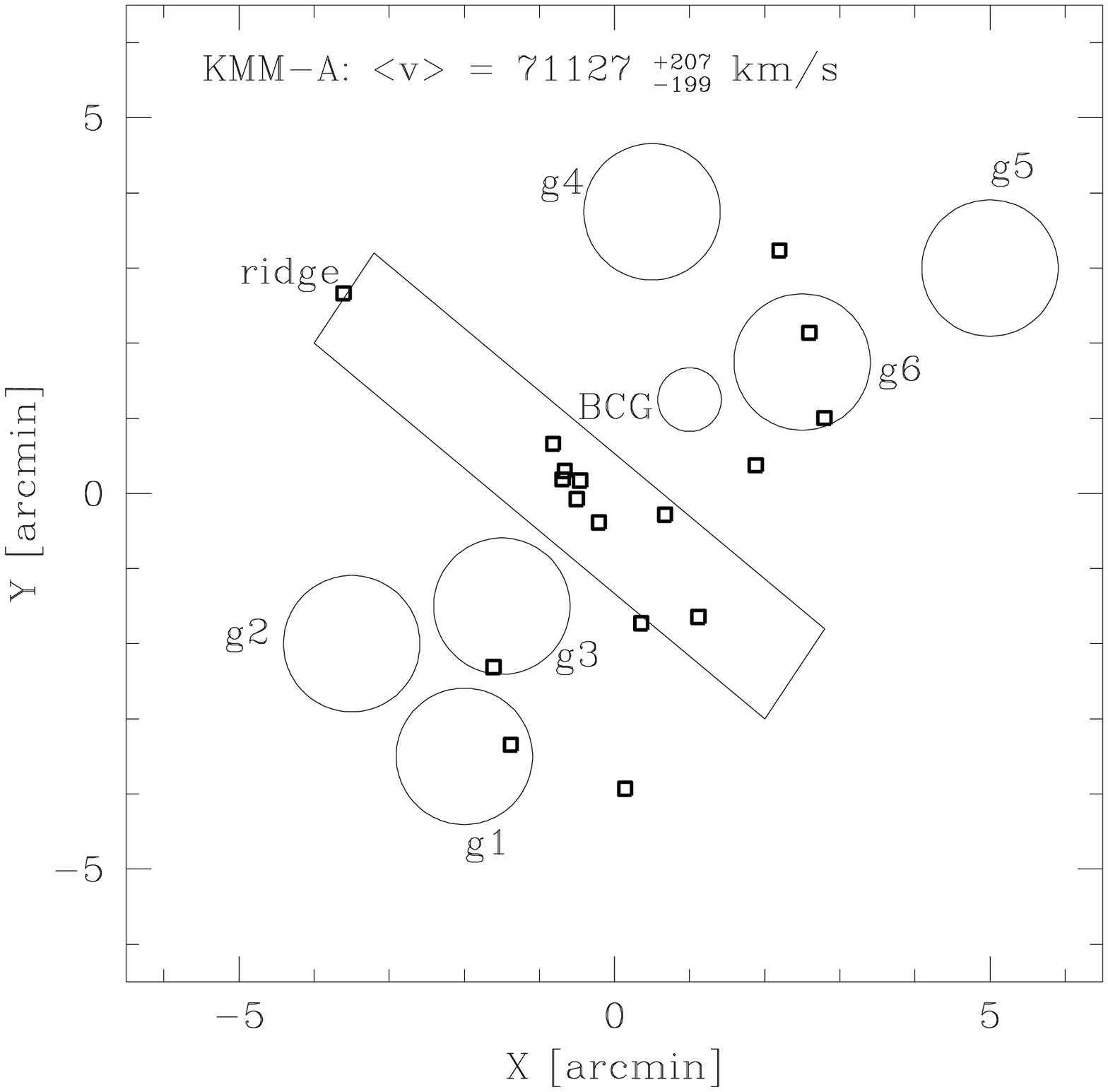}}

\resizebox{7.3cm}{!}{\includegraphics{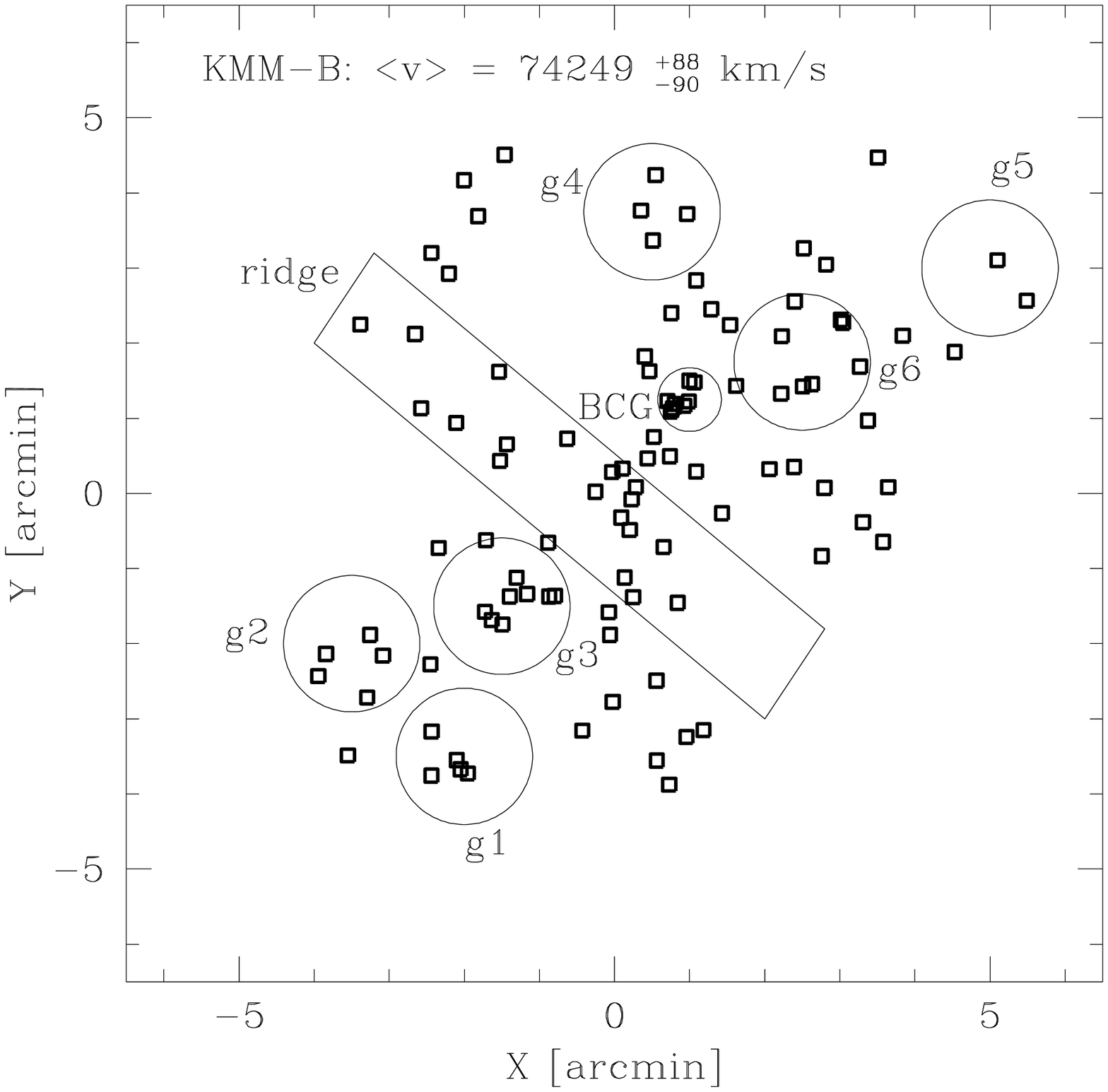}}

\resizebox{7.3cm}{!}{\includegraphics{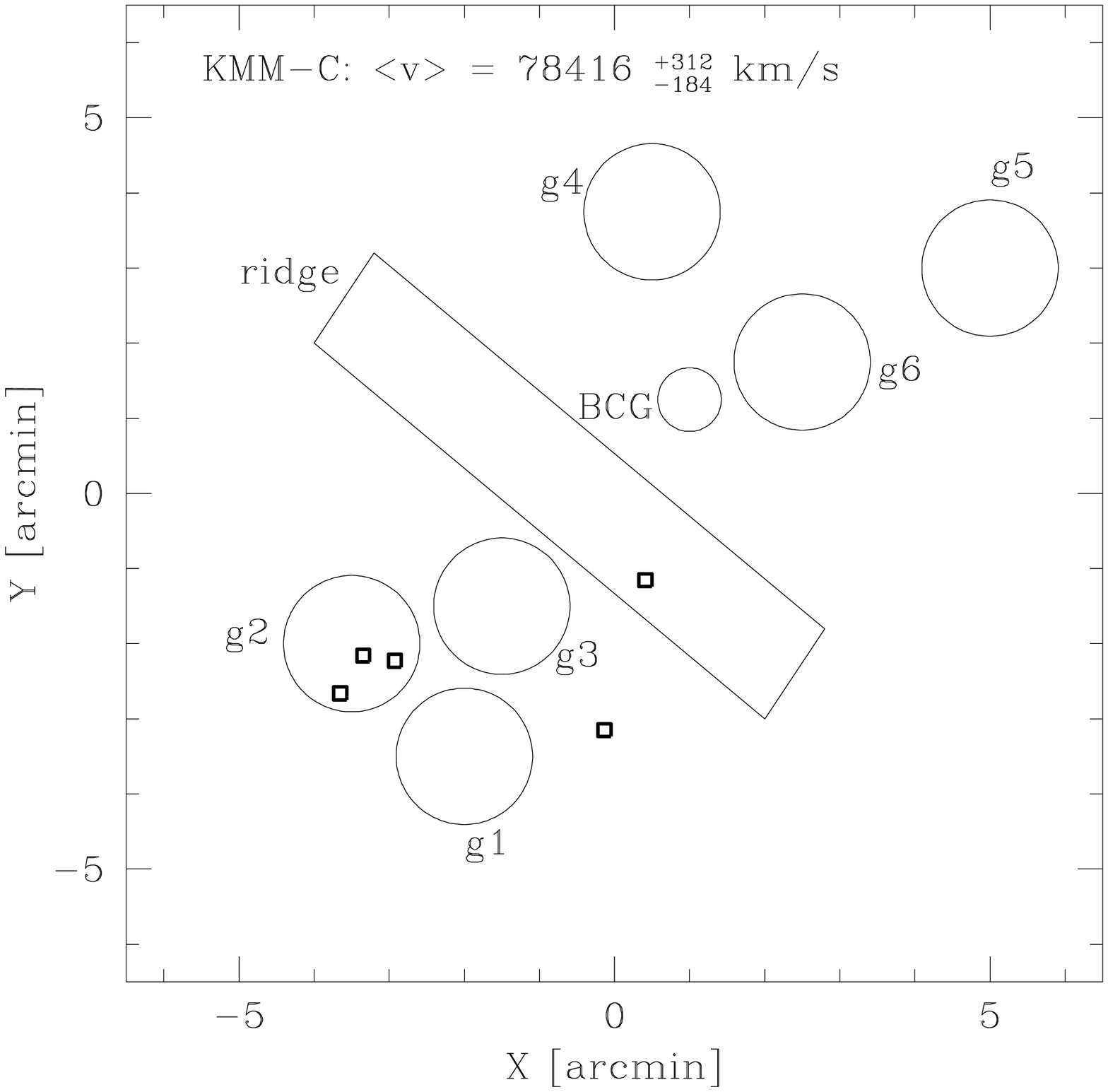}}
   
\parbox[b]{8cm}{
\caption[]{Projected coordinates of the galaxies assigned to the 
KMM partition. From top to bottom, KMM-A, KMM-B, KMM-C. The subgroups of the cluster identified by the isodensity contours in Figs.~\ref{FC} and~\ref{Slices} have been schematically represented (in next figures too).} 
\label{KMMxy}}  
\end{figure}  
  
We have also analyzed the higher moments of the distribution, 
in order to look  
for possible deviations from gaussianity that could provide important   
signature of dynamical processes. For all the following tests the null  
hypothesis is that the velocity distribution is a single Gaussian.  
The traditionally used shape estimators are kurtosis and skewness;  
in addition, we have computed the asymmetry and tail indices (AI and TI), 
which also measure the shape of a
distribution, but are based on the order statistics of the dataset
instead of its moments (Bird \& Beers, 1993).  
By definition, skewness, kurtosis and AI are equal to 0 for a gaussian
dish, while TI to 1.   
 In Table~\ref{subsTAB} we present the results; significance levels   
have been estimated from Table 2 in Bird \& Beers 1993.  
While the values obtained for skewness and AI  cannot allow to reject the   
Gaussian hypothesis (significance level~$>$~10\%), both kurtosis and TI 
indicate departure from a Gaussian distribution at better than 10\% 
significance level (Beers~\etal 1991). This indicates that the dataset has  
more weight in the tails than a Gaussian of the same dispersion.   
 
As departure from normality and high values of velocity dispersion   
can result from a mixing of several velocity distributions of smaller  
velocity dispersion with different locations, we have investigated  
various tests for the existence of substructure in the cluster velocity distribution.  
   
\begin{table}  
\begin{center}  
\caption{\label{subsTAB}{\small   
{\rm 1D substructure indicators for the 125 objects with quality   
flag=1 in our dataset}}}  
\begin{tabular}{ccc}  
\hline
\hline
Indicator & Value & Significance\\
\hline
AI & 0.470 & $\leq$0.20 \\  
TI & 1.113 & $\leq$0.10 \\  
Skewness & 0.087 & $>$0.20 \\  
Kurtosis & 0.779 & $\leq$0.05 \\ 
\hline  
\end{tabular}  
\end{center}  
\end{table}  
 
We have therefore addressed the presence of gaps which can be a 
signature of sub-clustering (Beers~\etal 1991). Five significant gaps  
in the ordered velocity dataset were detected. 
Fig.~\ref{stripe} shows the stripe density plots of radial
velocities of the 125 cluster galaxies and we have   
indicated the gap positions with an arrow, while in Table~\ref{GAP} one can   
find the velocity of the object preceding  the gap, the normalized size (i.e.,  
the ``importance'') of the gap itself, and the probability of finding a   
normalized gap of this size and with the same position in a normal   
distribution. Two very significant gaps are detected respectively at  
$\sim$ 76740 km/s (probability lower than 1.4\%)  
and $\sim$ 72150 km/s (probability lower than 0.2\%). 
   
\begin{table}  
\begin{center}  
\caption{\label{GAP}{\small   
{\rm Weighted gaps in the data}}}  
\begin{tabular}{ccc}  
\hline 
\hline  
Velocity [km/s] & Size & Significance \\
\hline
73455.4 & 2.298 & 0.030\\  
75243.4 & 2.323 & 0.030\\  
73704.0 & 2.454 & 0.030\\  
76738.1 & 2.577 & 0.014\\  
72150.0 & 3.089 & 0.002\\
\hline  
\end{tabular}  
\end{center}  
\end{table}

In addition, 9 of the 13 one-dimensional statistical tests of 
gaussianity performed by 
ROSTAT exclude the hypothesis of a single Gaussian distribution at better
than 10\% significance level (see Table~\ref{rostat_1Dstat}).
Among them, the Dip test is particularly indicative
(Hartigan \& Hartigan 1985). This tool tests the   
hypothesis that a sample is drawn from a unimodal, not necessarily  
Gaussian parent population. In the present case it rejects the unimodal 
hypothesis at a significance level better than 5\%, confirming our previous
results.

\begin{table}  
\begin{center}  
\caption{\label{rostat_1Dstat}{\small   
{\rm 1-D statistical tests performed in ROSTAT package that exclude the hypothesis of a single gaussian distribution. In Cols.~1 and 2 we report the name and the value of the statistics, while Col.~3 indicate their   
significance levels.}}}  
\begin{tabular}{ccr}  
\hline
\hline 
Statistical Test & Value & Significance\\
\hline
a & 0.743 & $\leq$0.10\\  
W & 0.969 & 0.08\\  
B2 & 3.779 & 0.04\\  
DIP & 0.023 & $\leq$0.05\\  
KS & 0.873 & 0.10\\  
V & 1.707 & 0.01\\  
${\rm W}^2$ & 0.162 & 0.02\\  
${\rm U}^2$ & 0.162 & 0.01\\  
${\rm A}^2$ & 0.989 &  0.01\\
\hline  
\end{tabular}  
\end{center}  
\end{table}

\subsection{Partitioning the distribution in velocity space}

In order to separate possible velocity groups within our velocity dataset  
we have used the KMM mixture modeling algorithm of McLachlan \& Basford   
(1988). This method has been shown to be very useful for detecting
  bimodality in astronomical datasets (Ashman and Bird 1994), and can
even be applied to detect multimodality.  One of the major
uncertainties however is the best choice of the  number of groups for the
partition. Given the appearance of the velocity histogram and of the
  strip density plot, and the presence of two highly significant gaps, we
  have chosen as a first guess to fit three velocity groups around the
  mean velocities 70000, 74000  and 78500 km/s. The KMM algorithm then
  fits a 3-group partition from this guess and optimizes the mean
  velocity for each group. However, in the case of a multimodal
  partition, the significance level is not determined accurately by
  the algorithm. The estimated P-value of 3 \% suggests a strong
  rejection of the null hypothesis (unimodal gaussian distribution)
  but has to be taken only as a guideline. 
The low-velocity  group A is best fitted by a Gaussian with parameters:  
${\rm C}_{\rm BI}=71127 ^{+207} _{-199} $~km/s and  ${\rm S}_{\rm BI}= 678 ^{+88} _{-69}$~km/s (17 objects).  
The major group B (103 objects) is found with    
${\rm C}_{\rm BI}=74249  ^{+88} _{-90}$~km/s  and  ${\rm S}_{\rm BI}=879 ^{+61} _{-55}$~km/s.   
At the high velocity tail, 5 galaxies  
are found, with location ${\rm C}_{\rm BI}=78416 ^{+312} _{-184} $~km/s, 
populating   
group C. The three gaussians corresponding to these partitions are displayed   
in Fig.~\ref{KMM}.   
  
We present in Fig.~\ref{KMMxy}   
the projected positions of the galaxies assigned to the three partitions.
The position of the main overdensities emerged in the projected density 
maps are also displayed for an easier reading.     
Galaxies in KMM-B are following the cross-like general pattern of the cluster. 
Half of the galaxies in KMM-A are located in the central ridge.
Galaxies belonging to 
KMM-C are all located in the South-East region of the cluster.
We have also tried partitions with a higher number of groups
(respectively 4 and 5) which also show strong rejection of the
 unimodal hypothesis. However,  as suggested by Ashman and Bird 1994, 
 we have followed the Occam's razor and selected the partition with the
smaller number of groups (tri-partition).

\section{Combined analysis of sub-clustering in velocity/2D space}

\subsection{Tomography of Abell 521}\label{VDP}

\begin{figure*}  
\centering
\resizebox{16cm}{!}{\includegraphics{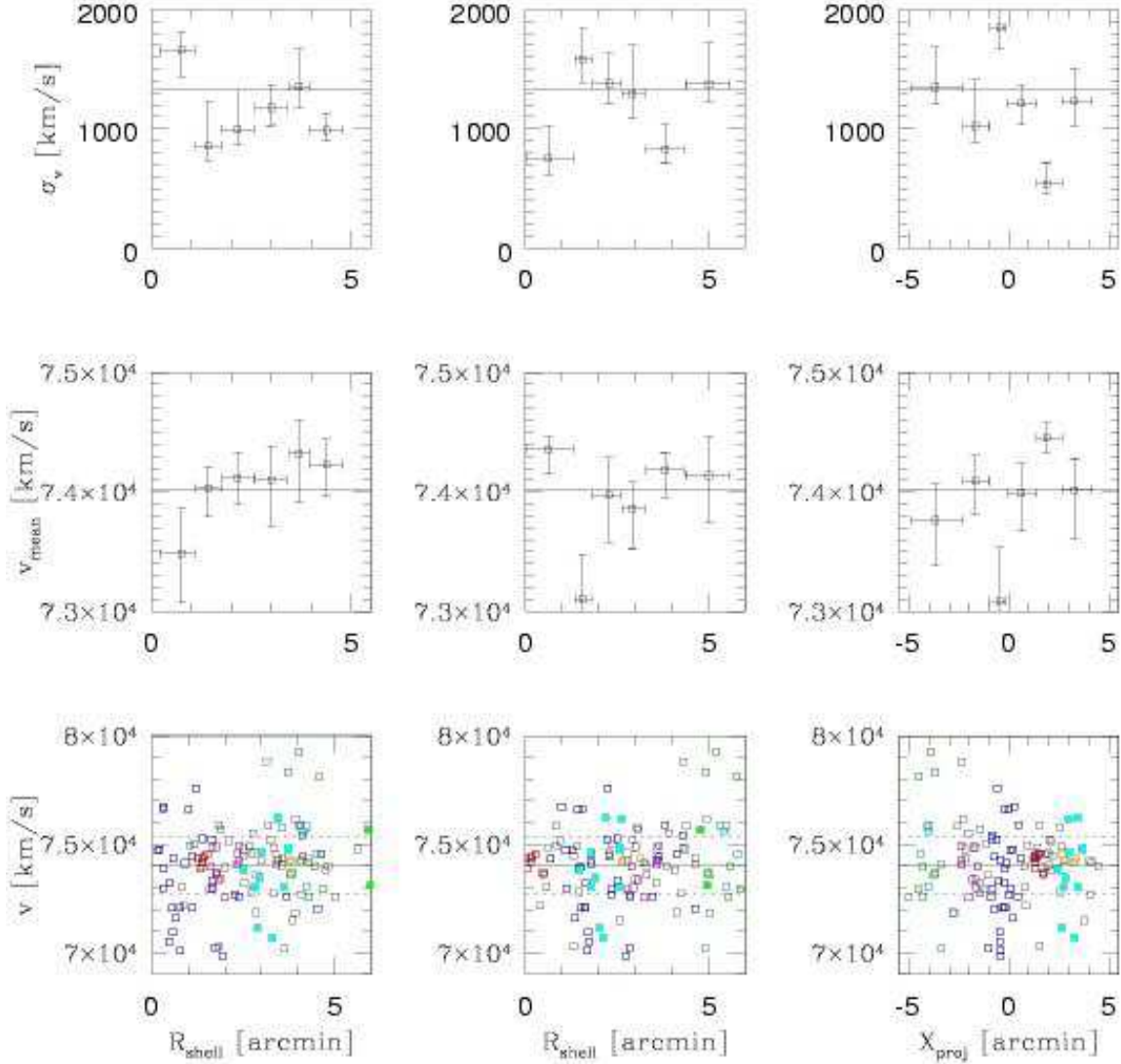}}  
\hfill  
\parbox[b]{16cm}{  
\caption{{\bf Left (from bottom to top):} radial velocities of the 125 galaxies of our spectroscopic catalogue, and velocity location and scale of concentric shells with a fixed number of objects (20), centered on the optical barycentre of the cluster. {\bf Center:} as before, but the new origin of the x-axis is the BCG position. {\bf Right:} as before, but ${\rm X}_{\rm proj}$ is the projected coordinate along the main axis of the cluster (NW/SE) and the center position is again on the optical barycentre of the cluster. Continuous lines represent the whole sample velocity scale (top) and location (mean and bottom). Dashed lines (bottom) indicate the [${\bar{\rm v}}-\sigma_{\rm v},{\bar{\rm v}}+\sigma_{\rm v}$] interval. Different colors correspond to the different subgroups detected on the isodensity map: 
{\bf red open squares:} BCG group~-~  
{\bf blue open squares:} ridge region~-~  
{\bf cyan open squares:} clump g1~-~ 
{\bf green open squares:} g2~-~ 
{\bf purple open squares:} g3~-~  
{\bf yellow open squares:} g4~-~
{\bf green full squares:} g5~-~
{\bf cyan full squares:} g6~-~
{\bf black open squares:} remaining objects}     
\label{sigprof}}  
\end{figure*} 

\begin{figure*}  
\resizebox{18cm}{!}{\includegraphics [angle=-90]{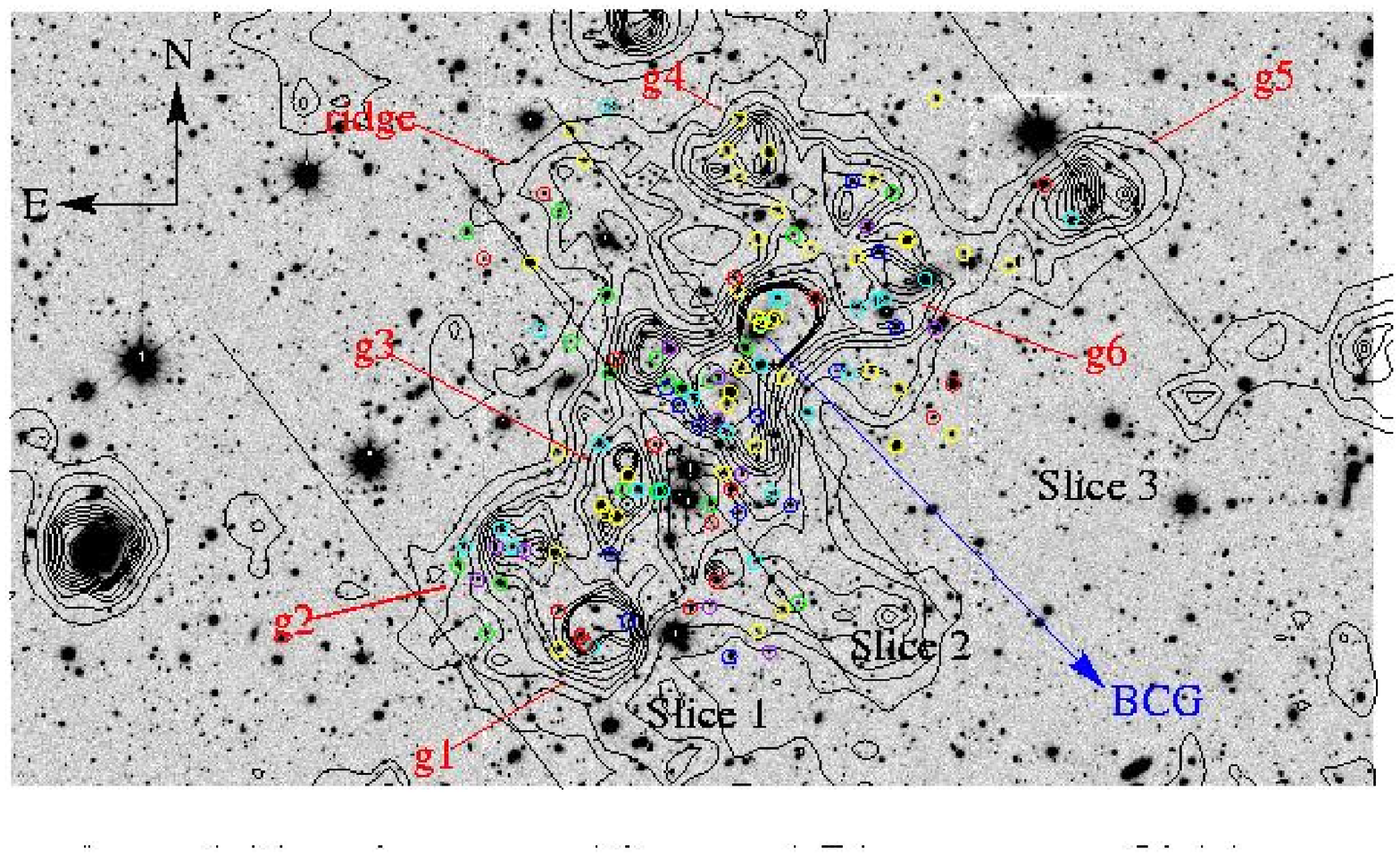}}   
\hfill  
\parbox[b]{18cm}{  
\caption{In black, galaxy iso-density contours for B$<$27 and I$<$20; the lowest iso-density contour corresponds to 1$\sigma$ level above the mean density in the field, the contours are spaced by 0.5$\sigma$. The division in 3 slices has been shown. Circles correspond to the galaxies in our high quality velocity dataset; different colors correspond to different ranges in radial velocity: 
{\bf blue:} ${\rm v}_{r}=69000{\div}72000$~-~  
{\bf green:} ${\rm v}_{r}=72000{\div}73000$~-~  
{\bf cyan:} ${\rm v}_{r}=73000{\div}74000$~-~ 
{\bf yellow:} ${\rm v}_{r}=74000{\div}75000$~-~ 
{\bf red:} ${\rm v}_{r}=75000{\div}76000$ 
{\bf purple:} ${\rm v}_{r}=76000{\div}80000$} 
\label{Slices}} 
\end{figure*}  

As indicated by the moments analysis, the tails of the velocity
distribution appear to be widely populated, suggesting 
the presence of interlopers. Moreover, the analysis of the number 
density maps of the cluster (Arnaud~\etal 2000 and Fig.~\ref{FC} of
the present paper) shows a complex structure, with six  groups
in projected coordinates (g1 to g6) along the main SW/NE direction 
of the cluster, a group around the BCG and a high density ridge in 
the direction perpendicular to the main axis of the cluster.

In Fig.~\ref{sigprof}, the values of the radial velocities 
(bottom), the mean radial velocities (middle), and the velocity dispersions
(top) are plotted as a function of an angular radius; velocity location and 
scale have been computed
in concentric shells with a fixed number of 
objects (20), in order to have comparable statistics. 
Different colors in the velocity vs. radius plots have been used to
visualize the different groups identified on the
isodensity map.

For the left panels, the center
is taken at the position of the main X-ray cluster (Arnaud~\etal 2000),
which roughly coincides with the barycentre of the galaxy distribution.
For the plots of the central column the origin is centered on the BCG
position, while in the the right panels we used the projected 
coordinate along the cluster main axis S2 (North-West/South-East).
Negative values correspond to the South-East extremity of the 
cluster, zero to the center, and positive values to the North-West
extent.

In Fig.~\ref{Slices}, galaxies have been circled
with different colors corresponding to different velocity bins.
Fig.~\ref{sigprof} and Fig.~\ref{Slices} can then be analyzed
together to understand the variation of the velocity distribution in
the field. 
The galaxies belonging to the region of the ``ridge'' S1,  
centered on the barycentre 
position and extending perpendicularly to the main axis S2, are 
color-encoded as blue open squares in Fig.~\ref{sigprof}. For this
particular region, we obtain systematically
 a lower mean velocity and a higher dispersion than for the whole cluster. 
This is particularly apparent in
Fig.~\ref{sigprof} top and medium, right panels.
This region consists of several clumps in projected density
map, but galaxies in the whole 
velocity range [70000-78000]~km/s populate the various clumps. 
However, it is the large number of low velocity objects that 
lowers the value of location in this region as compared to that
obtained for the whole cluster.

At $\sim$~1.5~arcmin from the barycentre of the cluster, in the 
North-East direction,  we find a compact region of galaxies at 
similar velocities corresponding to the BCG region 
(galaxies color-encoded as red squares in the bottom  panels
of Fig.~\ref{sigprof}), which 
shows a significantly lower velocity dispersion and a slightly
higher value of the mean velocity 
(top and medium panels of Fig.~\ref{sigprof}).

A high velocity 
group of four galaxies ($\sim$~78500~km/s) is detected at large
radius ($\sim$~4~arcmin) from the barycentre of the cluster 
(bottom  panel of Fig.~\ref{sigprof}), 
with a gap in velocity of more 
than 1500~km/s as compared to others galaxies at the same radius. 
Thus, these objects are likely to be unbound to the cluster;
they  are located in the South-East extremity of the cluster (bottom, right
panel of Fig.~\ref{sigprof}), and three of them in the region defined as g2 (green open
squares in Fig.~\ref{sigprof}). This latter is the
result of the superimposition along the line of sight of these  three 
high velocity galaxies and a concentration of
objects in the velocity range ([72000-74000]~km/s).

Several other clumps have a more homogeneous velocity composition:  g1
(cyan open squares in Fig.~\ref{sigprof}), g3 (purple open
squares in Fig.~\ref{sigprof}, and g4 ( yellow
open squares in Fig.~\ref{sigprof}); 
g1 is mainly populated with galaxies in the [74000-76000]~km/s
velocity range, while the greatest part of objects in g3 have velocity
in the [74000-75000]~km/s bin.
In particular, as one can see comparing the first and fifth bins
of the central, top panels of Fig.~\ref{sigprof}, g3 has a mean
velocity
and velocity dispersion quite close to the value of the BCG group.
The  North-East g4 clump also shows a  mean 
velocity location comparable to the BCG region, and a very small value
of velocity dispersion.  

Unfortunately, the number of measured radial velocities within the 
previous mentioned substructures is not always large enough to obtain 
meaningful dynamical information for each of them; we have thus divided
the cluster into three regions with a number of objects sufficient to 
derive stable estimators of location and scale and to get enough statistics 
in velocity histograms without degrading too much the binning.
These regions have been defined as three slices perpendicular to
the main direction of the cluster (see Fig.~\ref{Slices}). This choice 
has been motivated by the following reasons: first of all we wish to
test separately the velocity distribution within the high density ridge
S1 perpendicular to the direction of the main cluster S2, which was
shown in previous analysis to be of specific interest.
We therefore design the central slice (2) to include it.
We also want to investigate the possible differences in 
velocity distribution between the North-West and South-East regions
suggested in Fig.~\ref{sigprof}. The southern slice (1) 
includes the groups g1, g2 and g3; the northern slices (3) includes
both the BCG region as well as the northern extensions embedding
g4, g5 and g6 subgroups. In Fig.~\ref{plotN} the velocity 
distributions for each slice are shown; the corresponding values 
for velocity location and scale have been reported in Table~\ref{IdivTAB}.

In the southern region (slice 1), the mean velocity of the main structure,
obtained when excluding the four high velocity galaxies, is 
${\rm C}_{\rm BI}~=~73886 ^{+186} _{185}$~km/s, and its scale is
${\rm S}_{\rm BI}~=~1117 ^{+150} _{-86}$~km/s. The bimodal appearance
of the main structure histogram motivated us to try a KMM partitioning 
of the distribution. A very good agreement was found 
(0.929 significance level) for a fit by two gaussians centered 
respectively at 73070~km/s and 75140~km/s, and with velocity dispersions
of ${\rm S}_{\rm BI}~=~570 ^{+89} _{-52}$~km/s and 
${\rm S}_{\rm BI}~=~498 ^{+87} _{-40}$~km/s. This indicates that
in the  region there is probably a mix of two kinematically 
distinct populations in addition to the high velocity group, and
reflects the previous mentioned difference in velocity distribution
within the clumps g1 and g3 with respect to g2.

In slice~2, corresponding to the ridge region, the velocity distribution 
shows a very dispersed boxy shape, with a velocity dispersion reaching the
very large value of $1780 ^{+234} _{-142}$~km/s, in good agreement with
previous values by Maurogordato~\etal 2000. In agreement with
previous results, the location of the central slice is the lowest one
(${\rm C}_{\rm BI}~=~73625 ^{+344} _{-350}$~km/s).

The northern region (slice~3) shows a higher location than the previous
slices, ${\rm C}_{\rm BI}~=~74300 ^{+110} _{-104}$~km/s, which is 
comparable to the velocity of the BCG ($74357{\pm}44$~km/s),
and a lower velocity dispersion (${\rm S}_{\rm BI}~=~839 ^{+216} _{-134}$~km/s).
The shape of the distribution is symmetrical, although few galaxies are still
present in the low velocity tail. We have addressed the dynamics of the region
immediately surrounding the BCG; this galaxy has a complex structure, with a 
series of bright knots embedded in a lower density arclike structure at 
24~\h~kpc of its center (Maurogordato~\etal 1996, Maurogordato~\etal 2000). We 
have measured the redshifts of the four bright knots in the arclike
structure (see Table~\ref{tabBlobs}), 
showing that these objects belong to the cluster 
and are not gravitationally lensed background objects. The central
location in a $\sim$~240~\h~kpc region around the BCG (ten objects with
measured velocities including the BCG and its multiple nuclei) is 
${\rm C}_{\rm BI}~=~74340 ^{+40} _{-102}$~km/s. This values is very
close to the BCG radial velocity (74357$\pm$44~km/s); the velocity
scale is very low, $256 ^{+82} _{-133}$~km/s, and of the same order as 
the BCG internal velocity dispersion, 368$\pm$463~km/s 
(Maurogordato~\etal 2000) strongly suggesting it is bound  to the BCG.

When comparing the velocity distribution of the three slices, 
the northern region clearly shows a higher location and a lower velocity 
dispersion than the other regions, while a very high velocity dispersion
is observed within the central ridge. These results confirm the
trends that emerged in the velocity profiles in Fig.~\ref{sigprof}.

\begin{table} 
\begin{center} 
\caption{\label{IdivTAB}{\small  
{\rm Velocity distribution properties of the various subsamples of A521 galaxies }}} 
\begin{tabular}{cccc} 
\hline
\hline    
Subsample & Galaxy nb. & ${\rm C}_{\rm BI}$ & ${\rm S}_{\rm BI}$ \\
 & & [km/s] &  [km/s] \\
\hline
\multicolumn{4}{c}{ }\\  
Whole sample & 125 & $74019 ^{+112} _{-125}$ & $1325 ^{+145} _{-100}$   \\
\multicolumn{4}{c}{ }\\
KMM-A & 17 & $71127 ^{+207} _{-199}$ & $678 ^{+88} _{-69}$ \\
\multicolumn{4}{c}{ }\\
KMM-B & 103 & $74249  ^{+88} _{-90}$ & $879 ^{+61} _{-55}$ \\
\multicolumn{4}{c}{ }\\
KMM-C & 5 & $78416 ^{+312} _{-184} $ & - \\
\multicolumn{4}{c}{ }\\
Non-emission line & 110 &  $74166 ^{+96} _{-112}$ & $1087 ^{+117} _{-88}$   \\
galaxies & & & \\
\multicolumn{4}{c}{ }\\
Emission line & 15 & $72390 ^{+727} _{-923}$ & $2250 ^{+752} _{-384}$  \\
galaxies & & & \\ 
\multicolumn{4}{c}{ }\\
Early & 72 & $ 74083 ^{+121} _{-145}$ & $ 1105 ^{+135} _{-95}$ \\
\multicolumn{4}{c}{ }\\
Late & 41 & $73402 ^{+258} _{-278}$ & $1723 ^{+272} _{-170}$\\
\multicolumn{4}{c}{ }\\
Bright & 9 & $74321 ^{+118} _{-193}$ & $491 ^{+265} _{-243}$  \\
(${\rm I}_{\rm AB}<{{\rm I}_{\rm AB}}^*$)  & & & \\
\multicolumn{4}{c}{ }\\
Intermediate & 76 & $74000 ^{+144} _{-164}$  & $1331 ^{+156} _{-150}$ \\ 
(${{\rm I}_{\rm AB}}^*<{\rm I}_{\rm AB}<{{\rm I}_{\rm AB}}^*$+2)   & & & \\
\multicolumn{4}{c}{ }\\
Faint & 28 & $73515 ^{+268} _{-311}$ & $1635 ^{+449} _{-232}$ \\ 
(${\rm I}_{\rm AB}>{{\rm I}_{\rm AB}}^*$+2)  & & & \\
\multicolumn{4}{c}{ }\\
Slice 1 & 44 & $74037 ^{+233} _{-256}$ & $1454 ^{+266} _{-180}$ \\
\multicolumn{4}{c}{ }\\
Slice 1 without  & 40 & $73886 ^{+186} _{-185}$ & $1117 ^{+150} _{-86}$ \\ 
the background group &  & & \\
\multicolumn{4}{c}{ }\\
Slice 2 & 30 & $73625 ^{+344} _{-350}$ & $1780 ^{+234} _{-142}$ \\
\multicolumn{4}{c}{ }\\
Slice 3 & 51 & $74298 ^{+110} _{-104}$ & $839 ^{+216} _{-134}$ \\
\multicolumn{4}{c}{ }\\
\hline
\end{tabular} 
\end{center} 
\end{table}

\begin{figure}  
\centering
\resizebox{7cm}{!}{\includegraphics {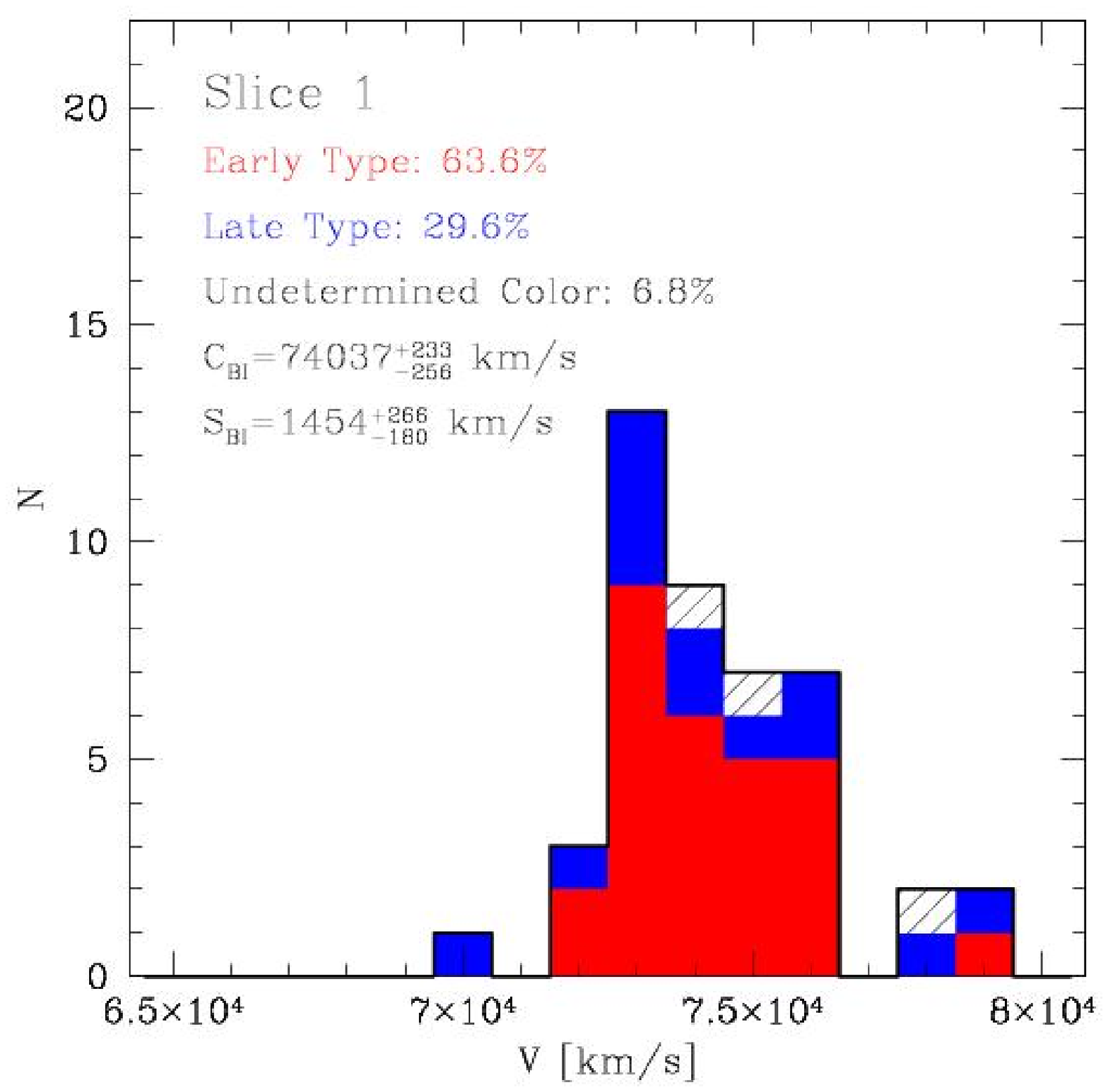}}
  
\resizebox{7cm}{!}{\includegraphics {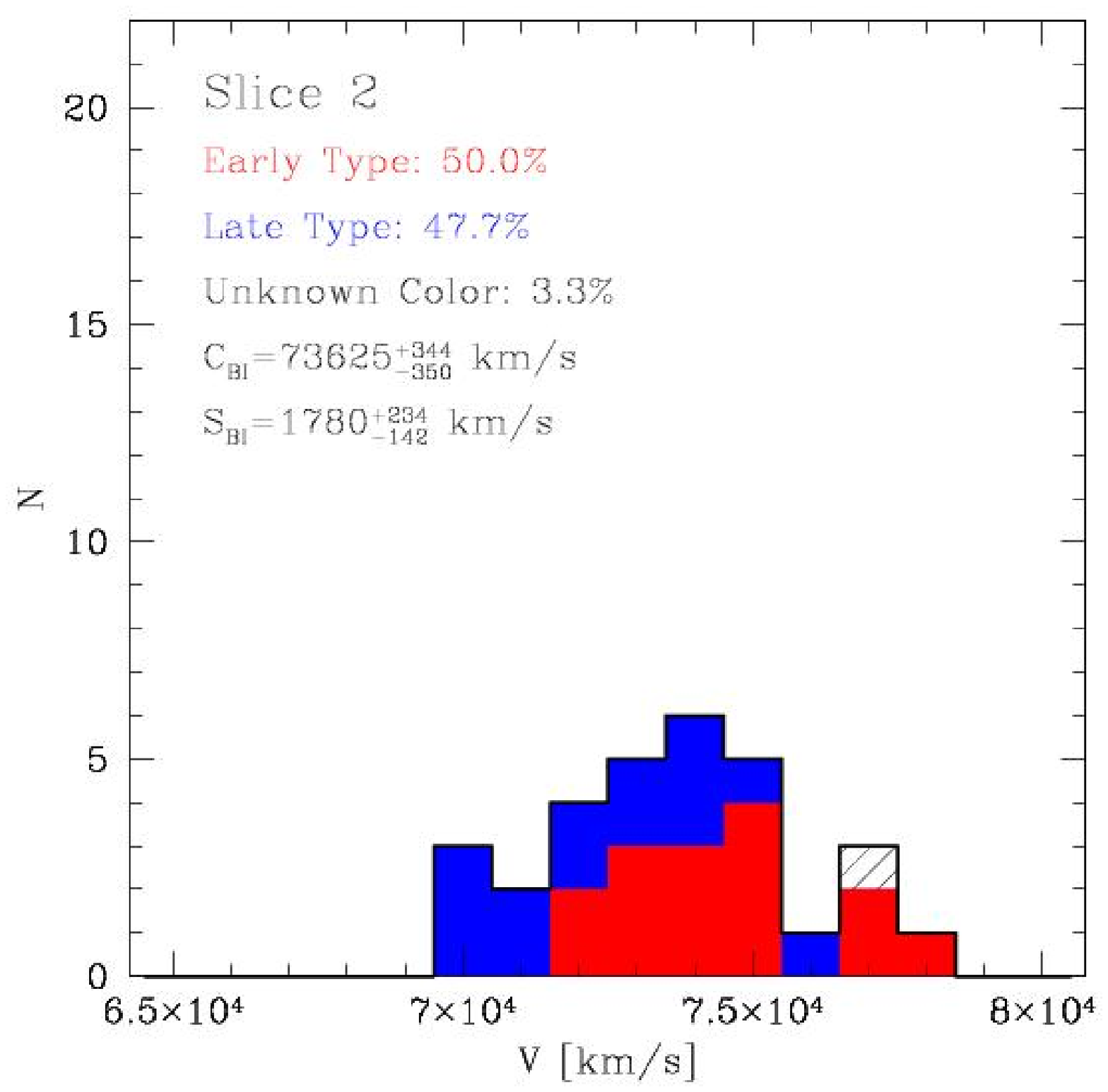}}

\resizebox{7cm}{!}{\includegraphics {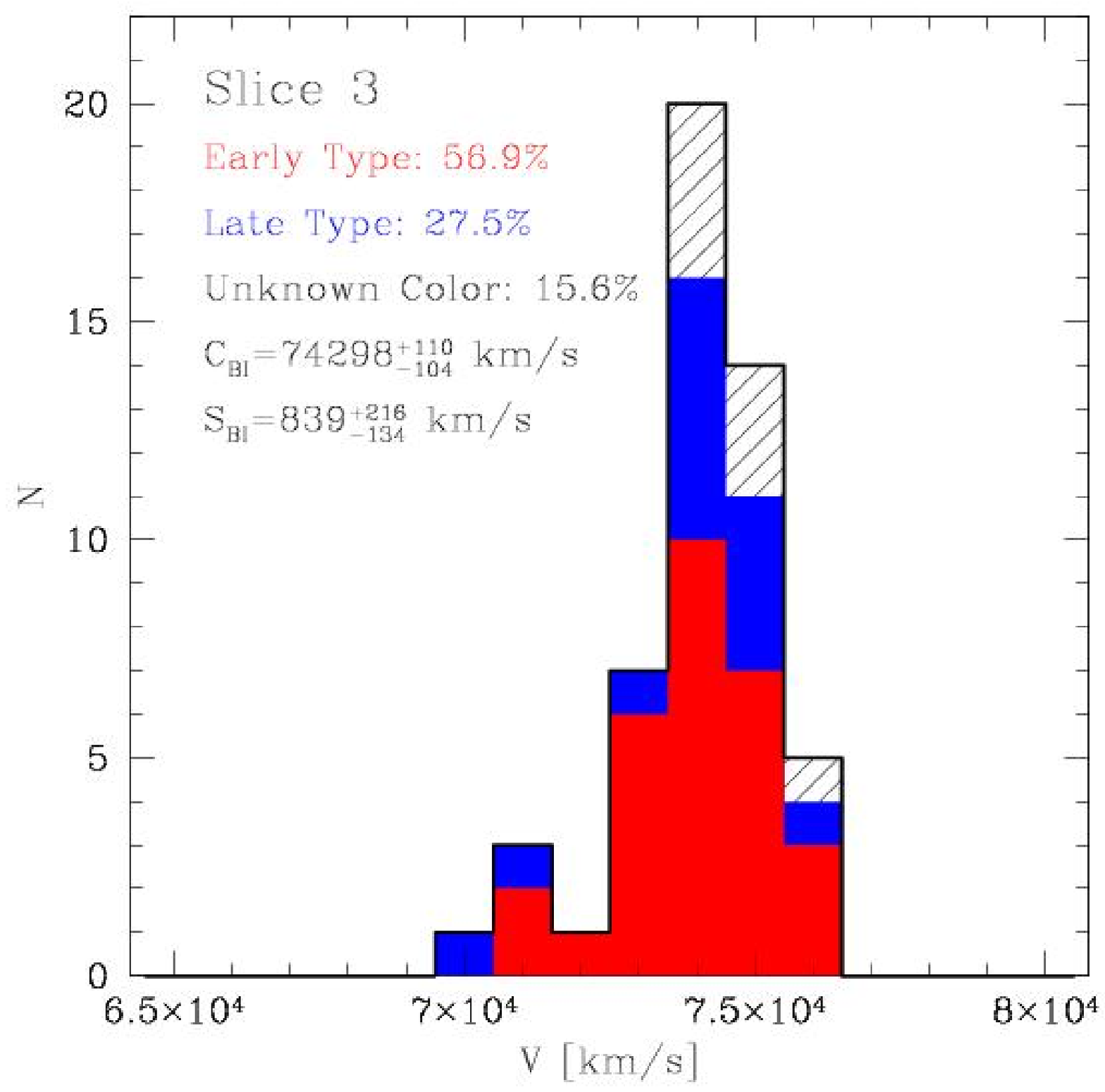}}  
\parbox[b]{8cm}{  
\caption{Velocity distributions of the three slices of Fig.~\ref{Slices}; the contribution due to different morphological types has been shown using different colors (early: red, late: blue, undetermined color objects: black shading). A binning of 1000~km/s has been used. Velocity location and scale found with ROSTAT are also shown.}  
\label{plotN}}  
\end{figure}  
 
\begin{table}  
\begin{center}  
\caption{\label{tabBlobs}{\small   
{\rm Radial velocities of the four blobs embedded in the arclike structure at 24~${{\rm h}_{75}}^{-1}$~kpc of the BCG center. The names of the blobs refer to Maurogordato~\etal 2000, Fig.~8.}}}  
\begin{tabular}{ccc}  
\hline  
\hline   
Blob & Radial Velocity(km/s) & Error on Velocity(km/s)\\ 
\hline  
B & 74340 & 100 \\ 
C & 74341 & 80 \\ 
D & 74325 & 52 \\ 
E & 74205 & 106 \\ 
\hline  
\end{tabular}  
\end{center}  
\end{table}

\begin{figure} 
\centering
\resizebox{6.5cm}{!}{\includegraphics  {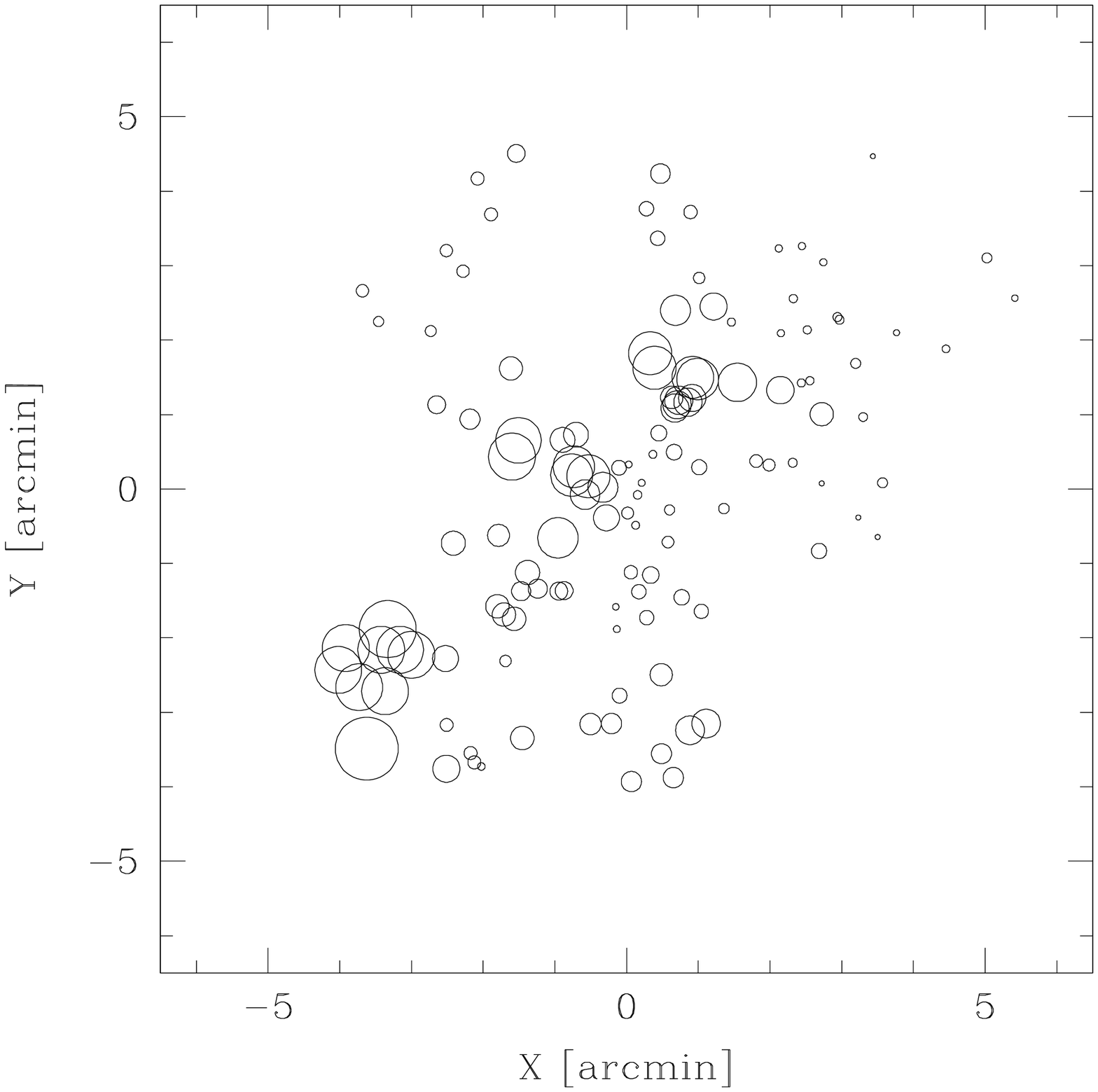}}

\resizebox{6.5cm}{!}{\includegraphics {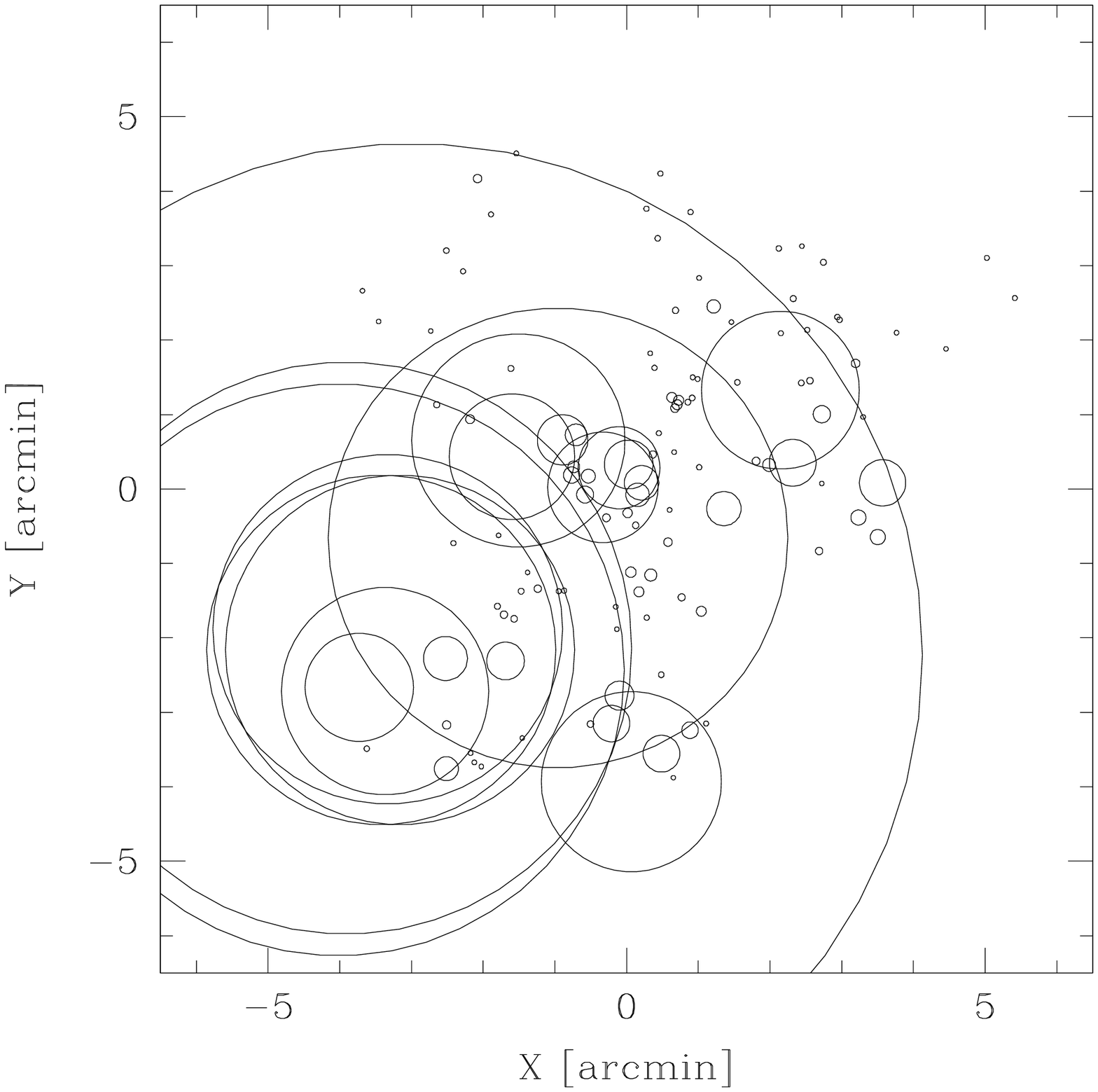}}

\resizebox{6.5cm}{!}{\includegraphics  {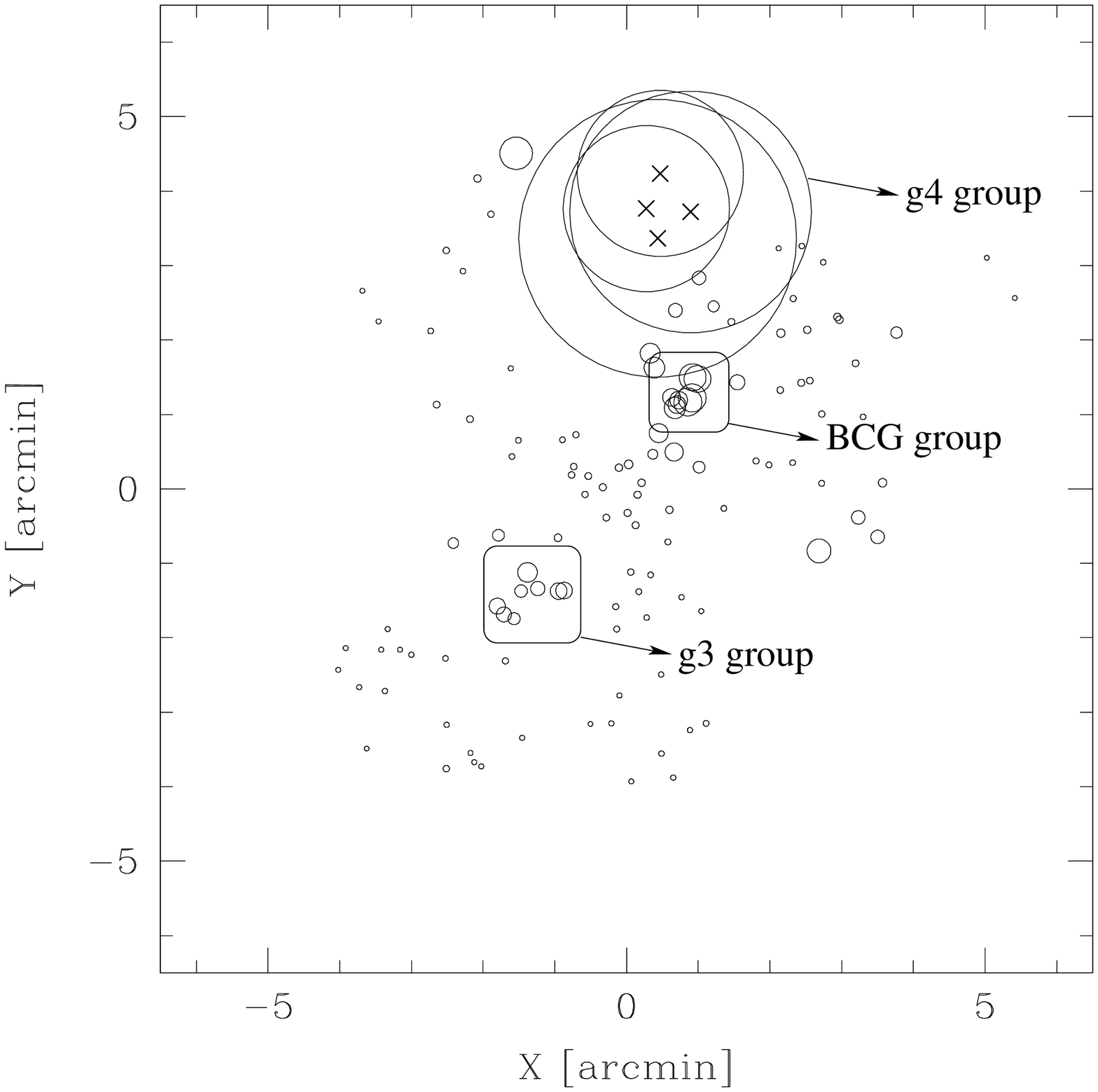}}

\parbox[b]{8cm}{  
\caption{Projected positions of the galaxies in our spectroscopic sample,
represented by circles with dimensions weighted by the estimators of Dressler
\& Shectman 1988 (${\Delta}$, top panel) and Girardi \etal~1997 
(${\delta}_{\rm V}$, middle panel , and ${\delta}_{\rm S}$,
 bottom panel). In the top panel, concentrations of large circles indicate a correlated spatial and kinematical variation. In the middle panel,  strong deviations of the local velocity as compared to
 the global one, while in the bottom panel, a low value of the local  velocity dispersion as compared to the mean one. Crosses correspond to circles larger than the box (see text for details).}
\label{Girardi}}  
\end{figure}

\subsection{Kinematical indicators of sub-clustering}  
  
Our previous analysis has clearly shown the  
presence of sub-clustering in the projected density distribution and 
the departure from gaussianity of the velocity distribution, 
indicating that the system has not yet reached 
equilibrium. As a second step, we have performed a more systematic search of 
sub-structures by addressing directly   
correlated deviations in position and velocity distributions.  
We have applied several classical methods that quantify the amount of   
substructures in galaxy clusters using positions and velocities.

In Table~\ref{test_3D} we list the actual values for $\Delta$ (Dressler \& Shectman, 1988), $\epsilon$ (Bird, 1994),  
and $\alpha$ (West \& Bothun, 1990) parameters and the significance of  
the corresponding tests, obtained through the bootstrap technique and 
by normalizing with 1000 Monte Carlo simulations.  
  
\begin{table}  
\begin{center}  
\caption{\label{test_3D}{\small   
{\rm 3-D substructure indicators for the sample of 125 objects with quality   
flag=1 in our dataset}}}  
\begin{tabular}{ccc}  
\hline 
\hline  
Indicator & Value & Significance\\  
\hline 
$\Delta$ & 164.8 & \bf {0.080}  \\  
$\epsilon$ & ${1.91}{\times}{10}^{+27}~{\rm kg}$ & 0.217 \\  
$\alpha$ & 0.183~${{\rm h}_{75}}^{-1}$~Mpc & 0.381 \\  
\hline  
\end{tabular}  
\end{center}  
\end{table}  
  
Assuming that these tests reject the null hypothesis if the significance
level is less than 10\%, only the $\Delta$~test finds evidences of 
subclustering at a high confidence level.
This result is further investigated by using the kinematical estimators 
introduced by Girardi~\etal (1997), which take into  
account separately the departures of the local mean (${\delta}_{\rm V}$) 
and dispersion (${\delta}_{\rm S}$)
from the global measurements for the whole cluster.
Low values of local velocity dispersion will give high values
of ${\delta}_{\rm S}$, while ${\delta}_{\rm V}$ will be high in case of strong 
departures (i.e. more than one $\sigma$) of the local mean velocity with 
respect to the global mean. 
In Fig.~\ref{Girardi}, each galaxy
is represented by a circle whose diameter is proportional to
$e^{{\Delta}}$ (top panel), $e^{{\delta}_{\rm V}} $ (middle panel) and $e^{{\delta}_{\rm S}}$ (bottom panel). The top panel shows several areas with many large circles, which indicate correlated spatial and kinematic variations. 
In the middle panel, very large circles are present in the South-East region of the cluster, corresponding to 
the group of background galaxies detected in previous sections. One can also note several large circles in the ridge region, corresponding to the presence of the previously detected low velocity group. 
Moreover, in the bottom panel, we detect three clumps in which the velocity 
dispersion is effectively low; one is in the South-East region
and corresponds to 
the structure identified as g3 in our iso-density maps, while the 
second is centered on the BCG complex.  
The third clump consists of a group of four galaxies in the g4 region;
it has such a low local velocity dispersion that 
the corresponding $e^{{\delta}_{\rm S}}$ values are larger than the 
box size of Fig.~\ref{Girardi}. Therefore, we have 
represented these objects with circles of diameter equal to their 
$\delta_{\rm S}$ parameters, and we have indicated their
projected positions with crosses.
Their radial velocities are in the range [74200-74500]~km/s, as can
clearly be seen in Fig.~\ref{sigprof}, where the four galaxies are 
represented by yellow squares. 
We also note that in the region of the ``ridge'' and in the extreme
South-East region, the dimension of the circles is the smallest, indicating in these regions 
the highest values of the velocity dispersion are reached.

\begin{figure*}  
\resizebox{18cm}{!}{\includegraphics{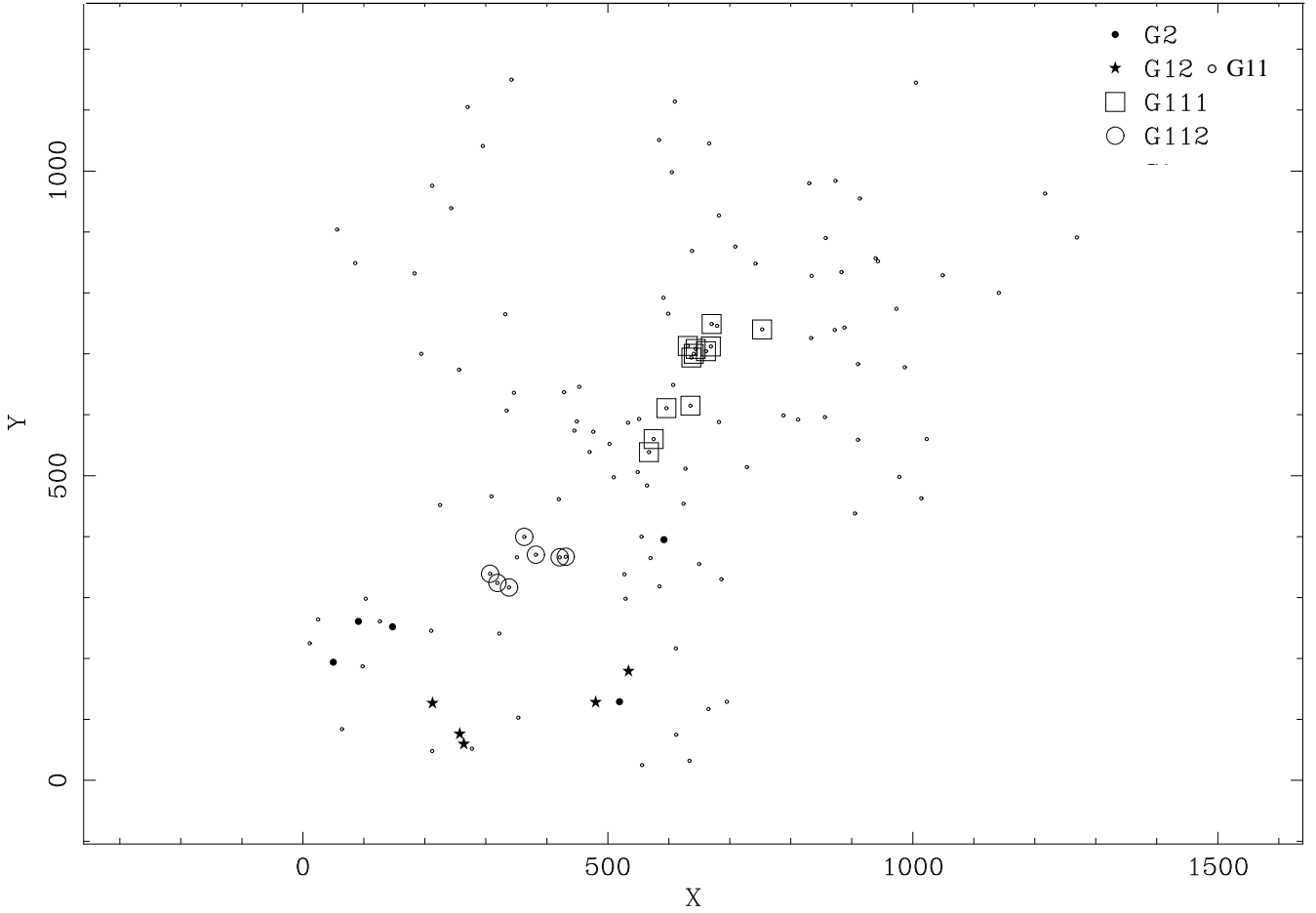}}  
\hfill  
\parbox[b]{18cm}{  
\caption{Substructure analysis by the h-tree method: 
the map in projected coordinates is shown, with different symbols 
corresponding to the different groups identified.}  
\label{htree2}}  
\end{figure*} 
   
We have also  applied the algorithm developed by Serna \& Gerbal 
(1996) which identifies subclusters on the basis of dynamical 
arguments. It uses the hierarchical clustering algorithm to  
associate galaxies according to their relative binding energies. 
Fig.~\ref{htree2} shows the map in projected coordinates visualizing
the groups resulting from the substructure analysis at the
various levels. At the first level, the algorithm separates 
the data into the main cluster G1 of 120 objects (open circles + stars) with 
${\rm v}_{mean}=73835$~km/s and $\sigma=1204$~km/s,  
and a high velocity set of 5 objects G2 (large filled points) in the
southern region, with  
${\rm v}_{mean}=78418$~km/s and  
$\sigma=  500$~km/s, previously identified as KMM-C. The algorithm
then splits again the main 
cluster  G1 in two components of very different mass ratios: the main
one G11 (93 objects, open circles) with a mean velocity similar to the value
obtained for G1 , but now a velocity dispersion
greatly reduced ($\sigma=940$~km/s), and a low mass group G12 of five
objects at higher velocity (75730 km/s)  plotted as stars in the 
southern region. These
objects were already detected in the analysis of
slice~1, as a possible higher velocity population of the southern
region. 
 At the last
level of structure identification, the algorithm finds out two
significant groups: the first one at North (G111), corresponding to the 
region surrounding  the BCG galaxy (squares), 
at a slightly higher mean velocity
than the mean component (74290 km/s) and presenting a low value of
the velocity dispersion (442 km/s), and a compact group southern of
the ridge (circles), corresponding to g3 in previous Sects.
The energy levels
of the various groups are also provided (not displayed): the deepest 
energy levels at the bottom of the energy well  
corresponds to the BCG complex. The 
lower level is occupied by the BCG itself associated to blob A, 
joining with the three other blobs C, D and E (see Table~\ref{tabBlobs})  
 at a slightly higher level.  
Associating to the BCG complex various galaxy pairs at low energy 
levels results into a bound  central group around the BCG galaxy.  
It is interesting to note that the BCG group and g3 are detected both
by the dynamical estimator of Girardi~\etal 1997, and by the 
${\rm h}_{\rm tree}$ algorithm.

\subsection{Dynamics of the groups}

As seen before, several groups are revealed from the sub-structure
analysis. We have then performed the gravitational bound check as in
the two-body problem (Beers, Geller and Huchra 1982):

\begin{equation}
{{\rm V}_r}^2{\rm R}_p~\leq~2~{\rm G}~{\rm M}~{\rm sin}^2{\alpha}~{\rm cos}{\alpha},
\end{equation}

\noindent where ${\rm V}_r$ is the relative line-of-sight velocity
of the two considered clumps, 
${\rm R}_p$ is the projected separation of the clump centers, 
$\alpha$ is the projected angle measured from the sky plane, and M is 
the total mass of the system.

In a first step, we have applied it to the groups G1 and G2 previously
defined by the h-tree method. 
The result is displayed in Fig.~\ref{unbound}, which shows
that the system is unbound for nearly all values of the projection angle
$\alpha$. We have then eliminated the group G2 from the analysis as a
background system of galaxies, and tested if the group G111
(corresponding to the system surrounding the BCG) and the group G112
(corresponding to g3) are bound to the remaining cluster. We
considered each time a two-body problem, with one system being the
tested group, and the other the remaining cluster. It results that the
groups G111 and G112 are  bound to the cluster for nearly all values of
the projection angle $\alpha$ (4 to 90 degrees).

\begin{table}  
\begin{center}  
\caption{\label{htreeTAB}{\small   
{\rm Properties of the significant groups from the ${\rm h}_{\rm tree}$ analysis}}}  
\begin{tabular}{ccccc}
\hline
\hline   
group &  ${\rm v}_{\rm mean}$ & $\sigma$ & ${\rm M}_{\rm vir}$ & ${\rm N}_{gal}$ \\
 & [km/s] & [km/s] & $(10^{15} M{\odot})$ & \\
\hline
main & 74018 & 1386 & 1.96  & 125 \\
G1 & 73835 & 1204  & 1.64  & 120 \\
G2 & 78418 & 504   & 0.2  &  5  \\
G11  & 73965 & 930   & 1.1 & 93\\
G12 & 75730 & 121 & 0.006 & 5 \\
G111 & 74290 & 442 & 0.045 & 12 \\
G112 & 74068 & 574 & 0.063 & 7 \\
\hline  
\end{tabular}  
\end{center}  
\end{table}

\begin{figure}  
\resizebox{8cm}{!}{\includegraphics {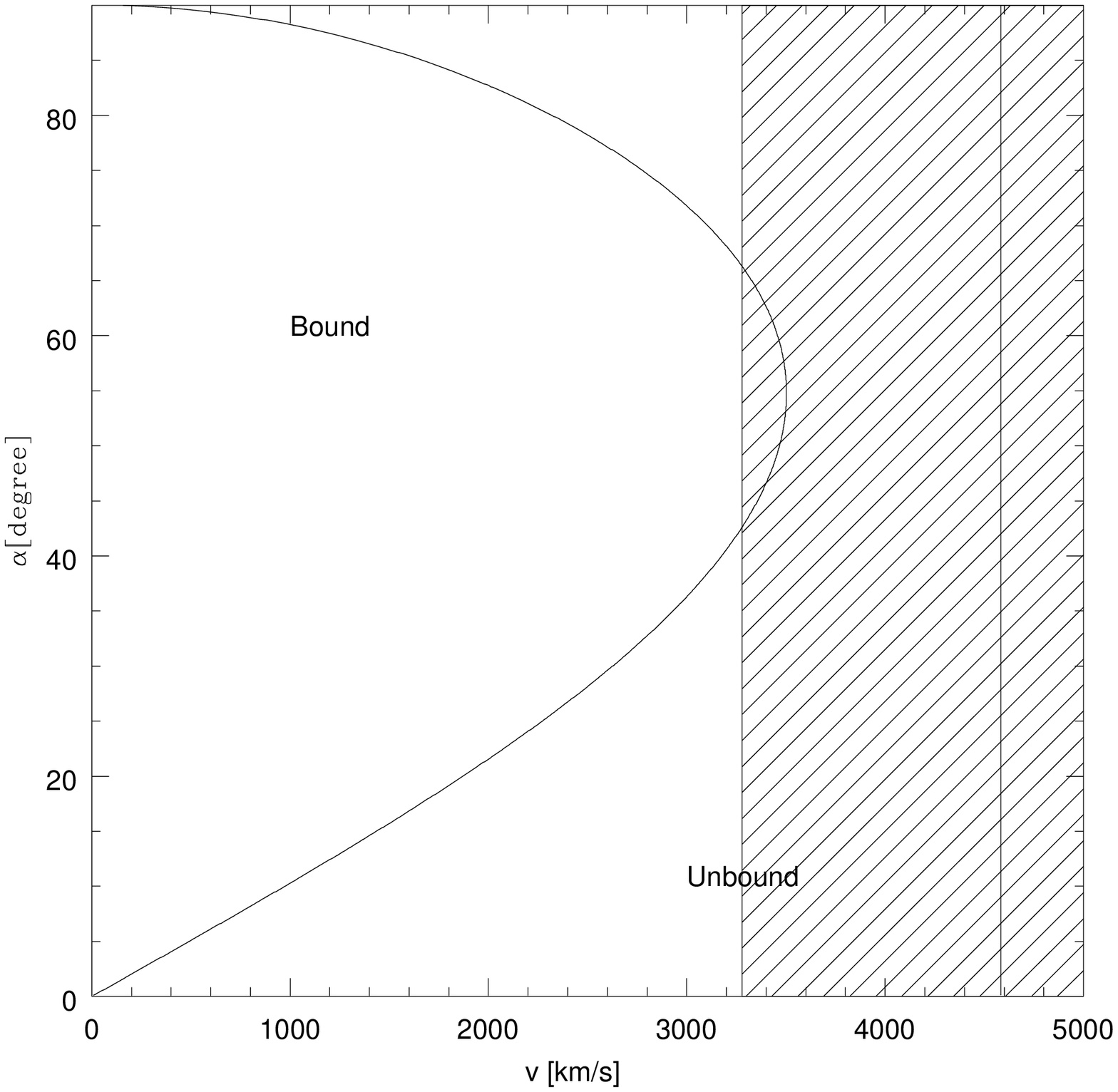}}  
\hfill  
\parbox[b]{8cm}{  
\caption{Boundary between the bound and unbound orbits for the main
cluster G1 and the high velocity system G2. The domain left of the
curve corresponds to bound orbits, while the right part to unbound
ones. The vertical line corresponds to the velocity difference between
the mean location of the two groups. The corresponding 68\% 
confidence region is shown 
with the cross-hatching.}  
\label{unbound}}  
\end{figure}

\section{Variation of dynamical properties with color and luminosity}\label{early_late}

In this Sect. we use the color information to define early and late
type galaxy subsamples, using the result that the bulk of early-type
galaxies in all rich clusters usually lie along a linear
color-magnitude relation. This so called ``red sequence'' has been
interpreted as a clear indication that all rich clusters contain a
core population of passively, evolving elliptical galaxies, coeval and
formed at high redshift (Ellis~\etal  1997, Kodama~\etal 1999, Gladders
\etal 1998). In Fig.~\ref{CMdiag} the (B-I) versus I
color-magnitude diagram (CMD) is shown for the objects in a field of
$15{\times}15~{\rm arcmin}^2$ 
(${\sim}3.25{\times}3.25~{{\rm h}_{75}}^{-2}~{\rm Mpc}^2$)
 centered on the optical barycentre of the cluster.
 
Due to the fact that we have no 
magnitude information for 12 of the 125 objects of A521, 
our velocity/magnitude catalogue is made up of 113 galaxies.

\begin{figure}  
\resizebox{8cm}{!}{\includegraphics{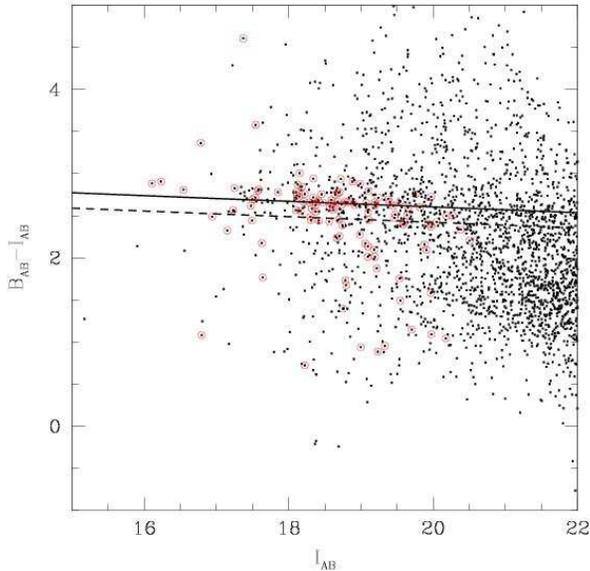}}  
\hfill  
\parbox[b]{8cm}{  
\caption{
The color-magnitude diagram (B-I) vs. I for all the galaxies in a
$15{\times}15~{\rm arcmin}^2$ area containing the core of the
cluster. The objects represented with a circle have spectroscopic
redshift and belong to the cluster. The solid straight line is the
linear best-fit found for the red sequence of the elliptical galaxies
of the cluster, while the dashed line represents the ``boundary''
between early and late type objects on the CMD.}
\label{CMdiag}} 
\end{figure}

The red sequence of the cluster has been characterized (slope,
intercept and width) by considering galaxies in a smaller field
($6.8{\times}6.8~{\rm arcmin}^2$, 
${\sim}1.5{\times}1.5~{{\rm h}_{75}}^{-2}{\rm Mpc}^2$) in
order to reduce the higher contamination in periphery by field objects.
We have used an
algorithm of linear regression plus an iterated 3$\sigma$ clip, following
and refining the method of Gladders~\etal  (1998) 
(detailed in Ferrari~\etal in prep.).  
The red sequence can then
be described by the linear equation 
${\rm (B-I)}_{\rm AB}=-0.033{\rm I}_{\rm AB}+3.273$
with a width of 0.183 (see Fig.~\ref{CMdiag}).  Galaxies with
measured redshift that lie within the identified red sequence 
and the three reddest objects of the cluster in the
color-magnitude diagram
have been classified as early type objects,
while the galaxies with bluer colors on this diagram have been assumed
to correspond to later types. 

We have analyzed the galaxy velocity distribution of our
sample as a function of this approximate early/late classification;
histograms of velocity and distribution in projected coordinates of
these objects are displayed in Fig.~\ref{Early-Late}.  Velocity
locations and scales of the two subsamples calculated using the
biweight technique are shown in Table~\ref{IdivTAB}.

\begin{figure}  
\resizebox{8cm}{!}{\includegraphics {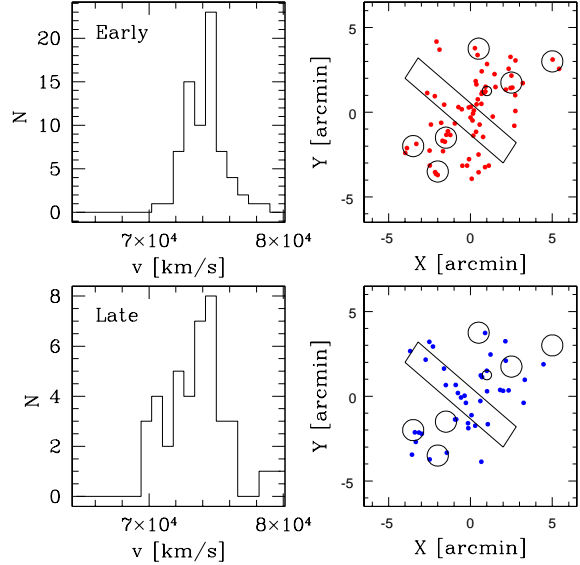}}  
\hfill  
\parbox[b]{8cm}{  
\caption{Top: Velocity distribution with a binning of 800~km/s (left) and projected coordinates (right) of early type galaxies. Bottom: as before, but for the late type subsample.}  
\label{Early-Late}}  
\end{figure}

We find significantly different velocity dispersions for the two
samples, with an extremely high value for the late type sub-sample,
while quite close to the value of the main cluster component (KMM-B)
for the early type one.

A trend in location is also detected, with the late type objects
located at a lower velocity as compared to the early type sample. A
Kolmogorov-Smirnov test rejects the hypothesis that the two velocity
datasets are drawn from the same parent population with a significance
level of $\sim$5.5\% (see table~\ref{K-S}).  The distribution in
projected coordinates is also very different for the two samples
(Fig.~\ref{Early-Late}, right panels). Most of the early type
galaxies are aligned with the S2 axis and populate the region of the
BCG and the groups, with some sparse objects in the outskirts, whereas
the galaxies of the late type sample are mainly located along
the S1 filament and in the external regions of the cluster, 
with very few objects in the South. This is confirmed by the velocity
histograms plotted in Fig.~\ref{plotN} for each slice of 
Fig.~\ref{Slices}, where
the counts in velocity bins are encoded with a different color for each 
morphological type (blue for late, red for early). 
The region of the ridge (slice 2) is the richest in late type objects 
($\sim50 \%$ versus $\sim 28 \%$ for slices 1 and 3), 
in particular in the low velocity tail.

We have then investigated if this strong morphological segregation could
 be related at least partially to a luminosity segregation. 
Using the previous classification, we have converted apparent
magnitudes to absolute ones, assuming k and e-corrections for E and Sa
models given by Poggianti (1997). We have then divided the catalogue
in three subsamples in absolute magnitudes, corresponding to ${\rm
I}_{\rm AB}{\leq}{{\rm I}_{\rm AB}}^*$, ${{\rm I}_{\rm AB}}^*<{\rm
I}_{\rm AB}<{{\rm I}_{\rm AB}}^{*}+2$, ${\rm I}_{\rm AB}{\geq}{{\rm
I}_{\rm AB}}^{*}+2$ (with ${{\rm I}_{\rm AB}}^{*}=-22.7$, derived
using the ${\rm i'}^*$ value in Goto~\etal 2002 and 
following the indication of Fukugita~\etal 1995 for transforming i' in 
${\rm I}_{\rm AB}$),
in order to test if the spatial and velocity distribution differs for
various luminosity classes within the cluster. In table~\ref{IdivTAB}
we report the results for each sub-sample.

In Fig.~\ref{Idiv}, we display the projected distributions and the  
velocity histograms for the three tested subsamples. 
As a first evidence, the velocity dispersion increases drastically from 
brightest to faintest objects, with more than a factor two between the  
value obtained for the ${\rm I}_{\rm AB}{\leq}{{\rm I}_{\rm AB}}^{*}$  
sample towards the ${\rm I}_{\rm AB}{\geq}{{\rm I}_{\rm AB}}^{*}+2$ one. 
There is also some indication of shift in location, as the brightest  
sub-sample is characterized by a mean velocity $\sim$700~km/s higher than  
the fainter ones.

\begin{figure}  
\resizebox{8cm}{!}{\includegraphics {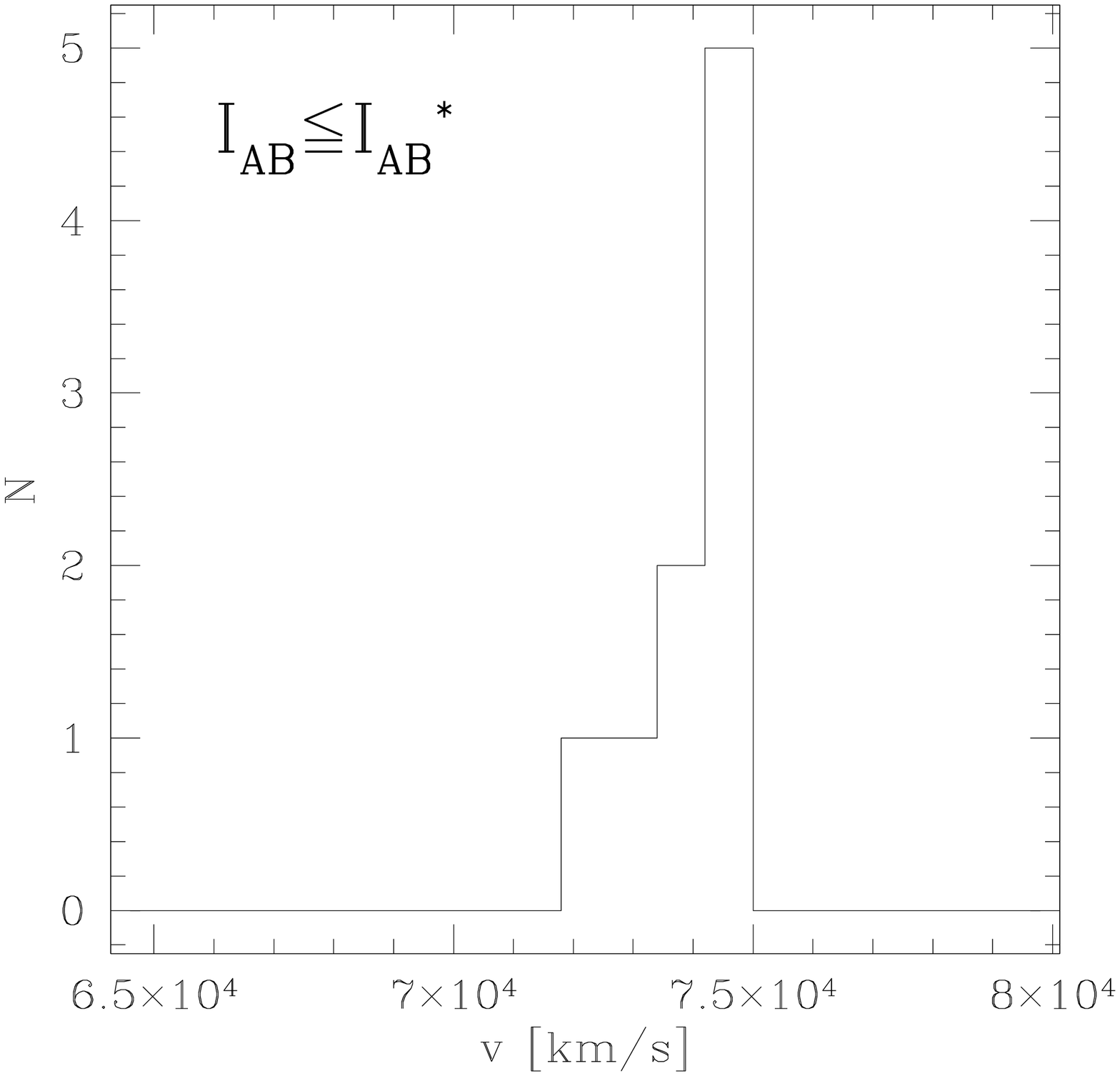}
\includegraphics {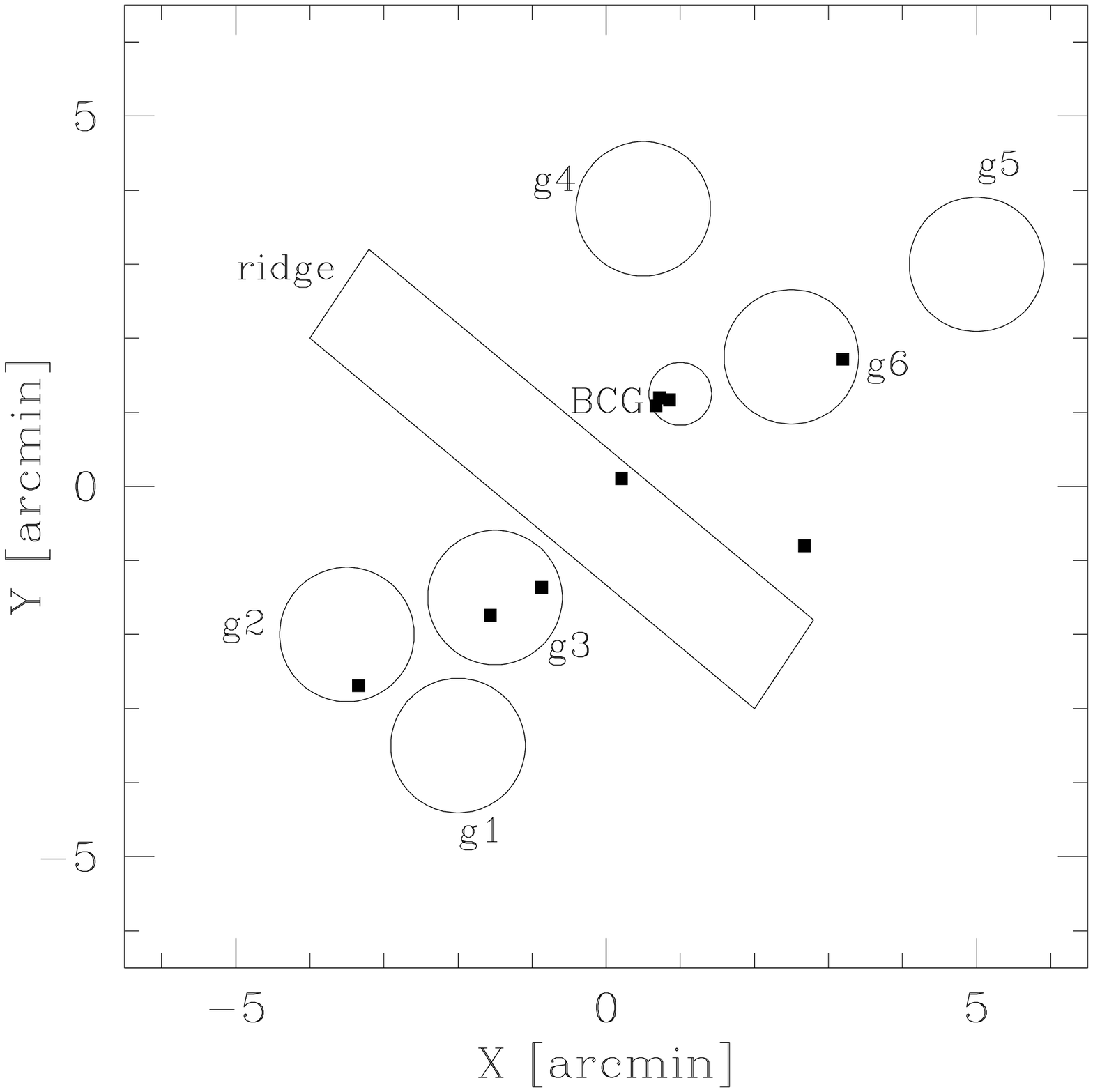}}
\hfill  
\resizebox{8cm}{!}{\includegraphics {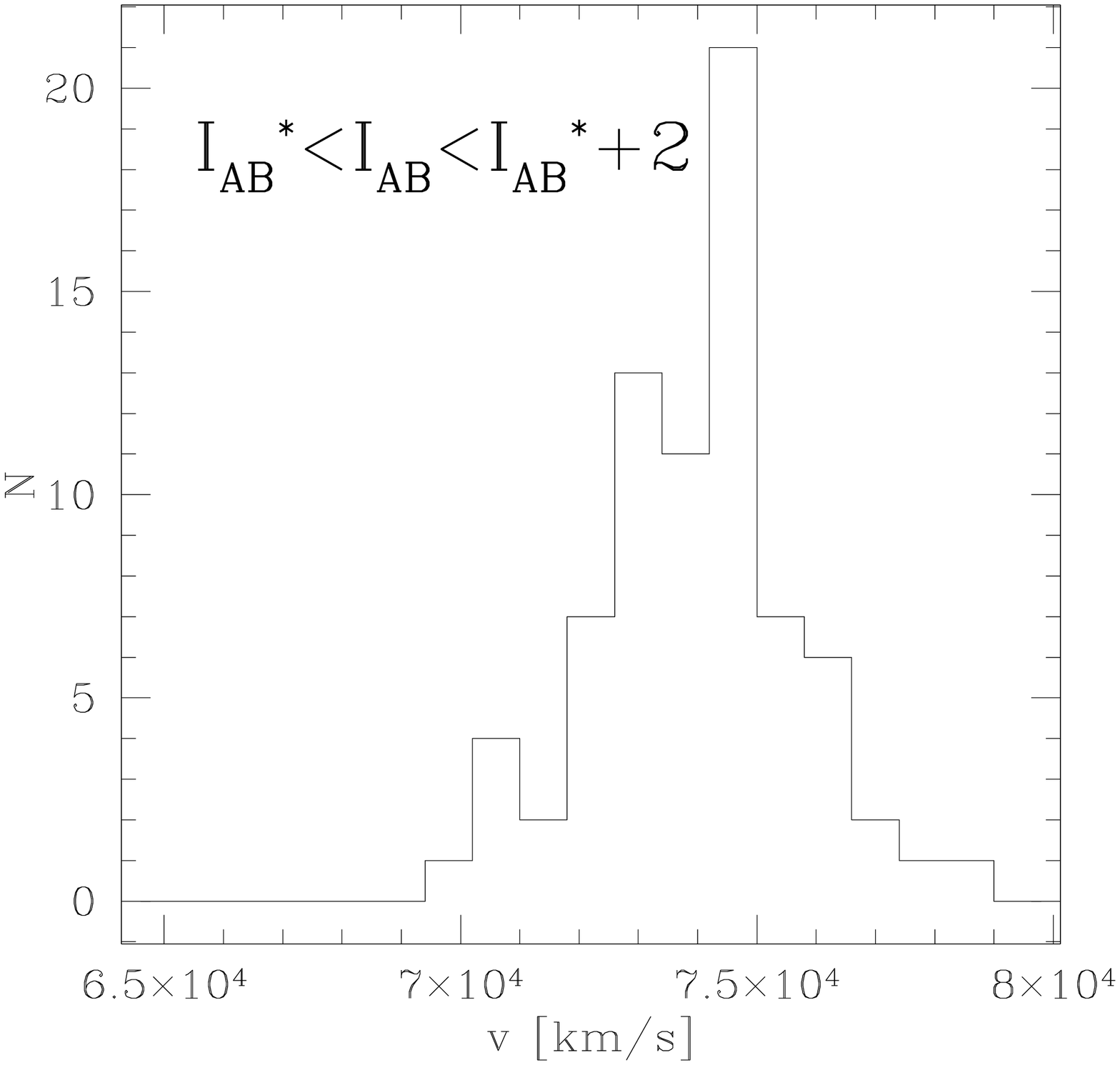}
\includegraphics {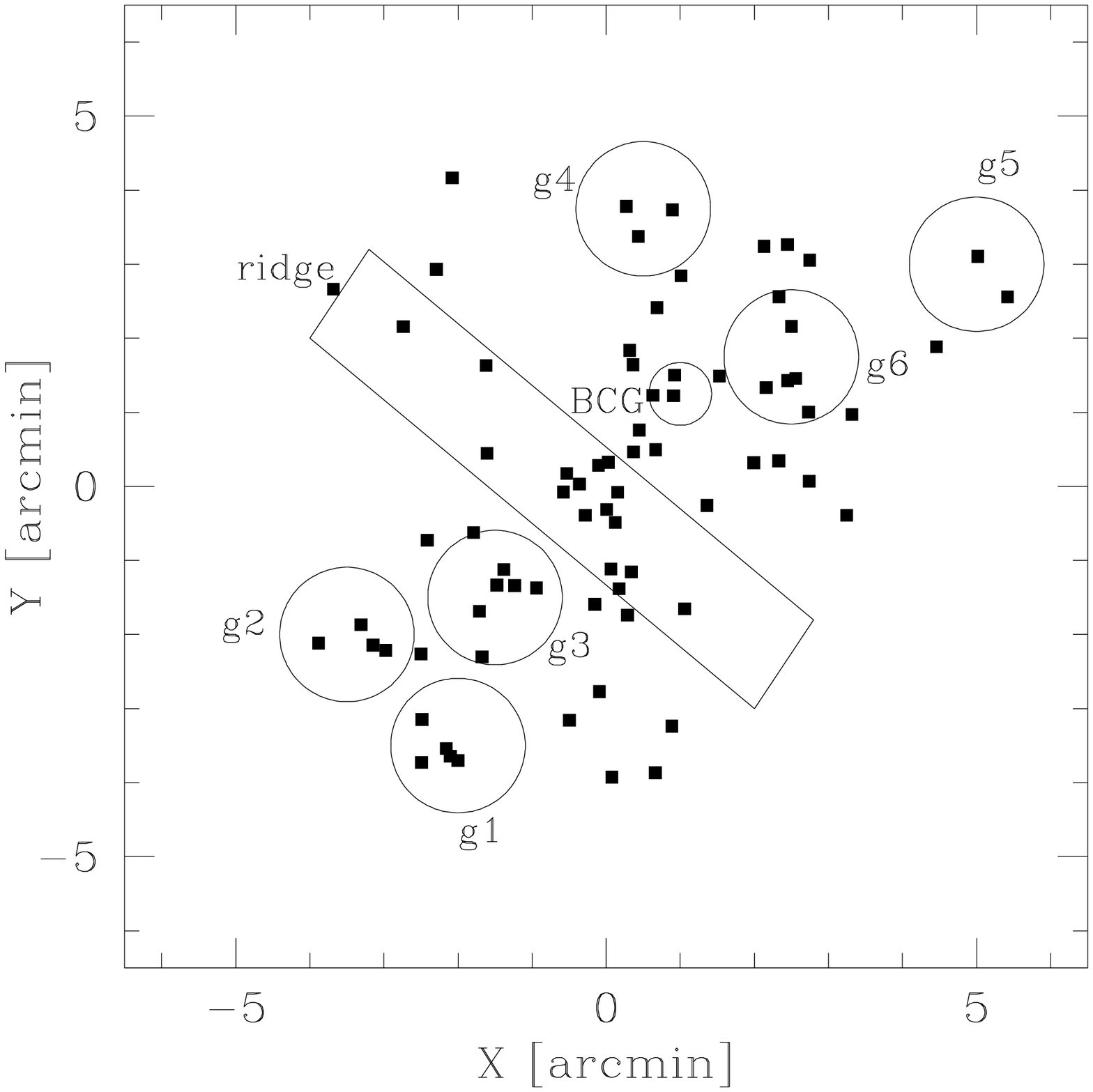}}
\hfill
\resizebox{8cm}{!}{\includegraphics {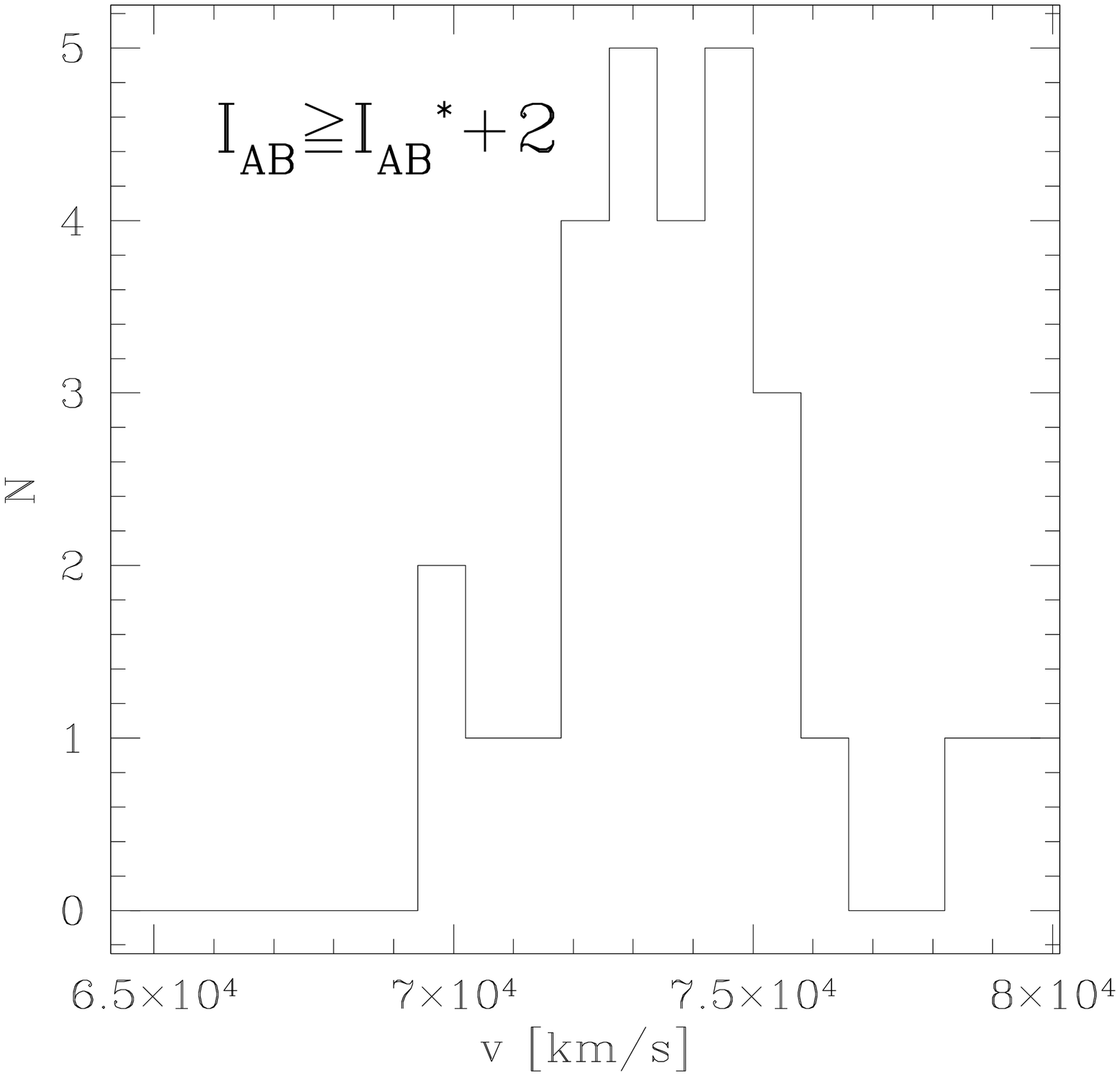}
\includegraphics {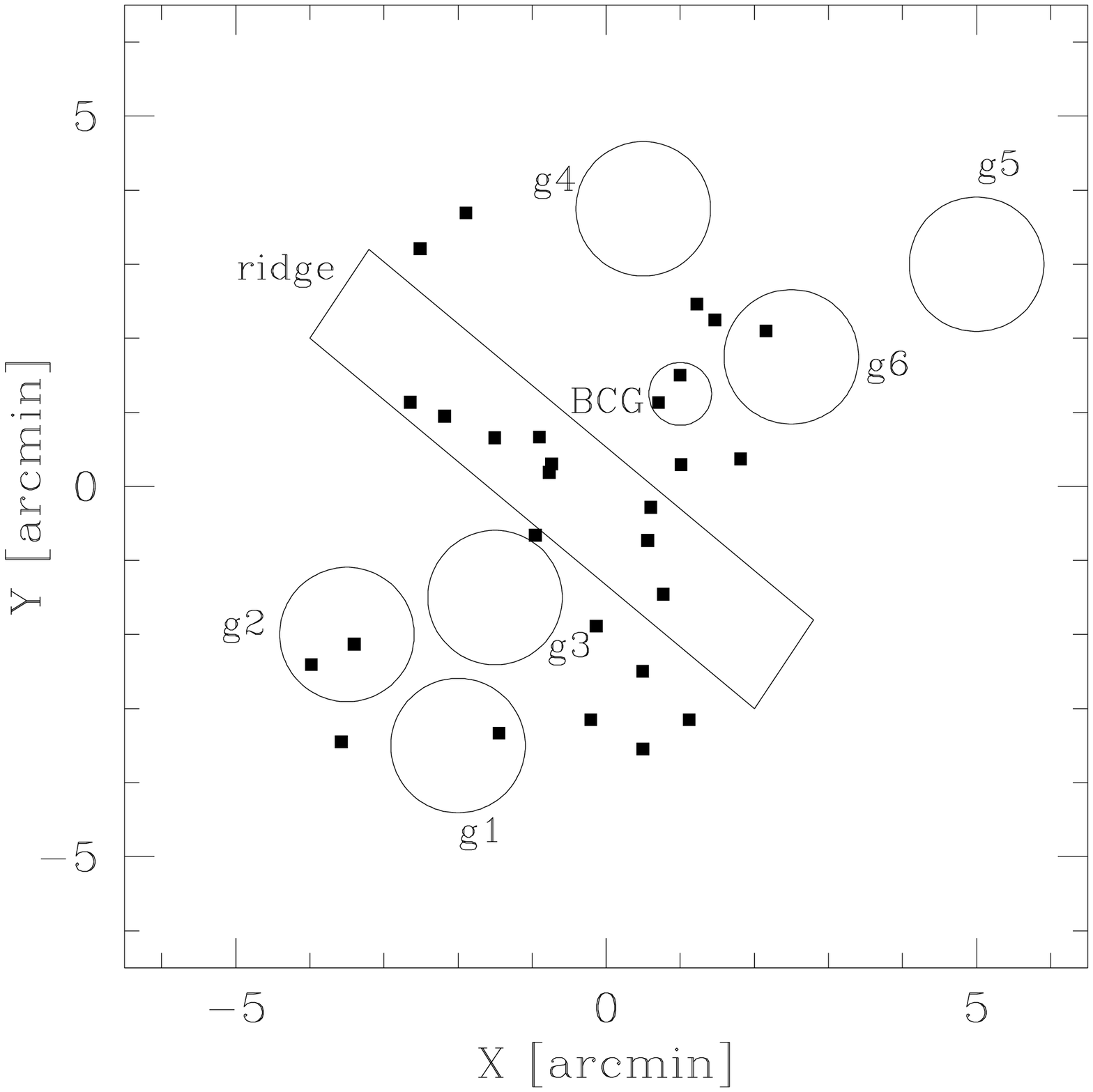}}
\hfill
\parbox[b]{8cm}{  
\caption{Spatial and velocity (with a binning of 800 km/s) distribution for the three ${\rm I}_{\rm AB}$ magnitude subsamples of A521 galaxies. Top: ${\rm I}_{\rm AB}{\leq}$-22.7 - Middle: -22.7$<{\rm I}_{\rm AB}<$-20.7 - Bottom: ${\rm I}_{\rm AB}{\geq}$-20.7}  
\label{Idiv}} 
\end{figure}

The brightest galaxies lie only on  the main axis of the cluster (S2), 
while the  faintest ones are preferentially located along S1 
with some galaxies in the outskirts.
We have then tested for correlations between the pseudo-spectral type
and luminosity class distributions. 
The late type subsample and the faint one  
have extremely similar distributions both in velocity and in 
projected positions;  
in both cases, a K-S test indicates  
that they are drawn from the same distribution function with  
high significance  
levels (more than 99\% and 94\%, Table~\ref{K-S}). 
On the other side, both the velocity and the projected position  
distributions of the early type and of the intermediately luminous  
objects are drawn from the same parent population with a significance level  
of more than 97\%, while the early type and the brightest galaxies have
significance levels higher than 80\%.  

\begin{table}   
\begin{center}
\caption{\label{K-S_Slices}{\small   
{\rm Significances (\%) given by the Kolmogorov-Smirnov Test for the velocity distributions of the three slices of Fig.~\ref{Slices}}}}  
\begin{tabular}{|c|c|c|c|}
\hline
 & & & \\
 & Slice 1 & Slice 2 & Slice 3 \\
 & & & \\
\hline
 & & & \\
Slice 1 & -- & 37.4 & 27.1 \\ 
 & & & \\
\hline
 & & & \\
Slice 2 & -- & -- & 12.3 \\
 & & & \\
\hline
\end{tabular}  
\end{center}  
\end{table}

\begin{table}   
\begin{center}
\caption{\label{K-S}{\small   
{\rm Significances given by the Kolmogorov-Smirnov Test for the velocity (top) and spatial (bottom) distributions of the various detected sub-samples}}}  
\begin{tabular}{|c|c|c|c|c|c|}
\hline
\multicolumn{6}{|c|}{ }\\
\multicolumn{6}{|c|}{Significances (\%)} \\
\multicolumn{6}{|c|}{for velocity distributions} \\
\multicolumn{6}{|c|}{ }\\
\hline
\hline
 & & & & & \\
 & Early & Late & Bright & Interm. & Faint \\
 & & & & & \\
\hline 
 & & & & & \\
Early & -- & 5.7 & 81.8 & \bf {97.9} & 22.2 \\ 
 & & & & & \\
\hline
 & & & & & \\
Late & -- & -- & 34.8 & 30.7 & \bf {99.7} \\
 & & & & & \\
\hline
 & & & & & \\
Bright & -- & -- & -- & 58.7 & 20.4 \\
 & & & & & \\
\hline
 & & & & & \\
Interm. & -- &  -- & -- & -- & 56.1 \\
 & & & & & \\
\hline 
\hline
\multicolumn{6}{|c|}{ }\\
\multicolumn{6}{|c|}{Significances (\%)} \\
\multicolumn{6}{|c|}{for projected position distributions} \\
\multicolumn{6}{|c|}{ }\\
\hline
\hline
 & & & & & \\
 & Early & Late & Bright & Interm. & Faint\\
 & & & & & \\
\hline
 & & & & & \\
Early & -- & 47.4 & 87.2 & \bf {99.9} & 69.8 \\ 
 & & & & & \\
\hline
 & & & & & \\
Late & -- & -- & 77.0 & 92.6 & \bf{94.2} \\ 
 & & & & & \\
\hline
 & & & & & \\
Bright& -- & -- & -- & 77.0 & 64.3 \\
 & & & & & \\
\hline
 & & & & & \\
Interm. & -- &  -- & -- & -- & 35.6 \\
 & & & & & \\
\hline 
\end{tabular}  
\end{center}  
\end{table}

We have also analyzed separately the population of emission lines (15 objects) 
versus non emission lines galaxies (110 objects) (Fig.~\ref{noemi-emi}). 
When excluding emission lines
galaxies, the location remains comparable but the
velocity dispersion is notably smaller than that of the whole sample 
(Table~\ref{IdivTAB}). 
As shown by the histogram in Fig.~\ref{noemi-emi}, the velocity distribution 
of emission lines galaxies consists of a major concentration at 
low velocity (around 70000 km/s) and several isolated objects spanning the 
whole range of the cluster.
Considering these objects together with the non-emission lines objects 
results in strongly enhancing the global velocity dispersion.

\begin{figure*}  

\resizebox{16cm}{!}{\includegraphics {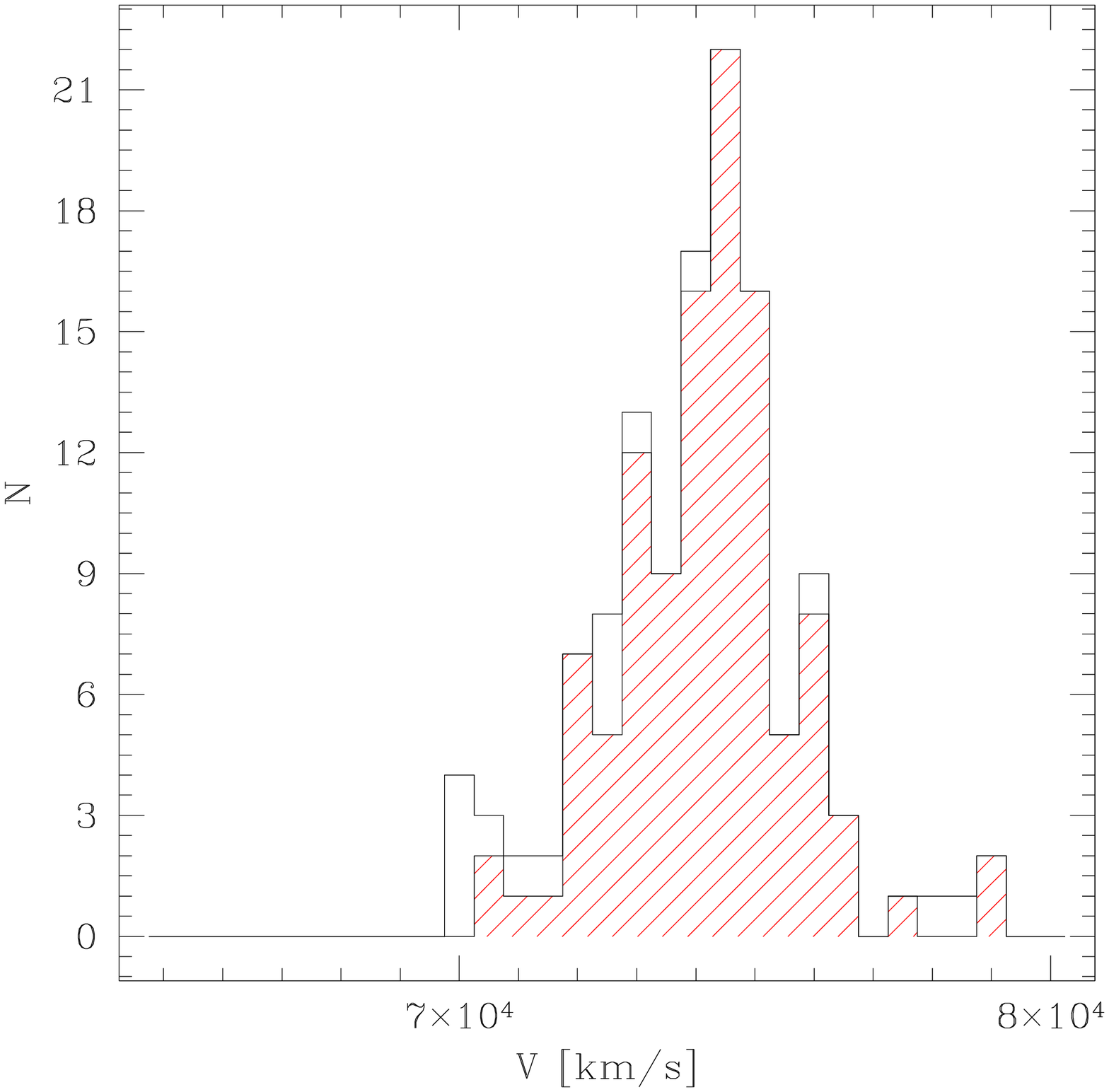}
\includegraphics {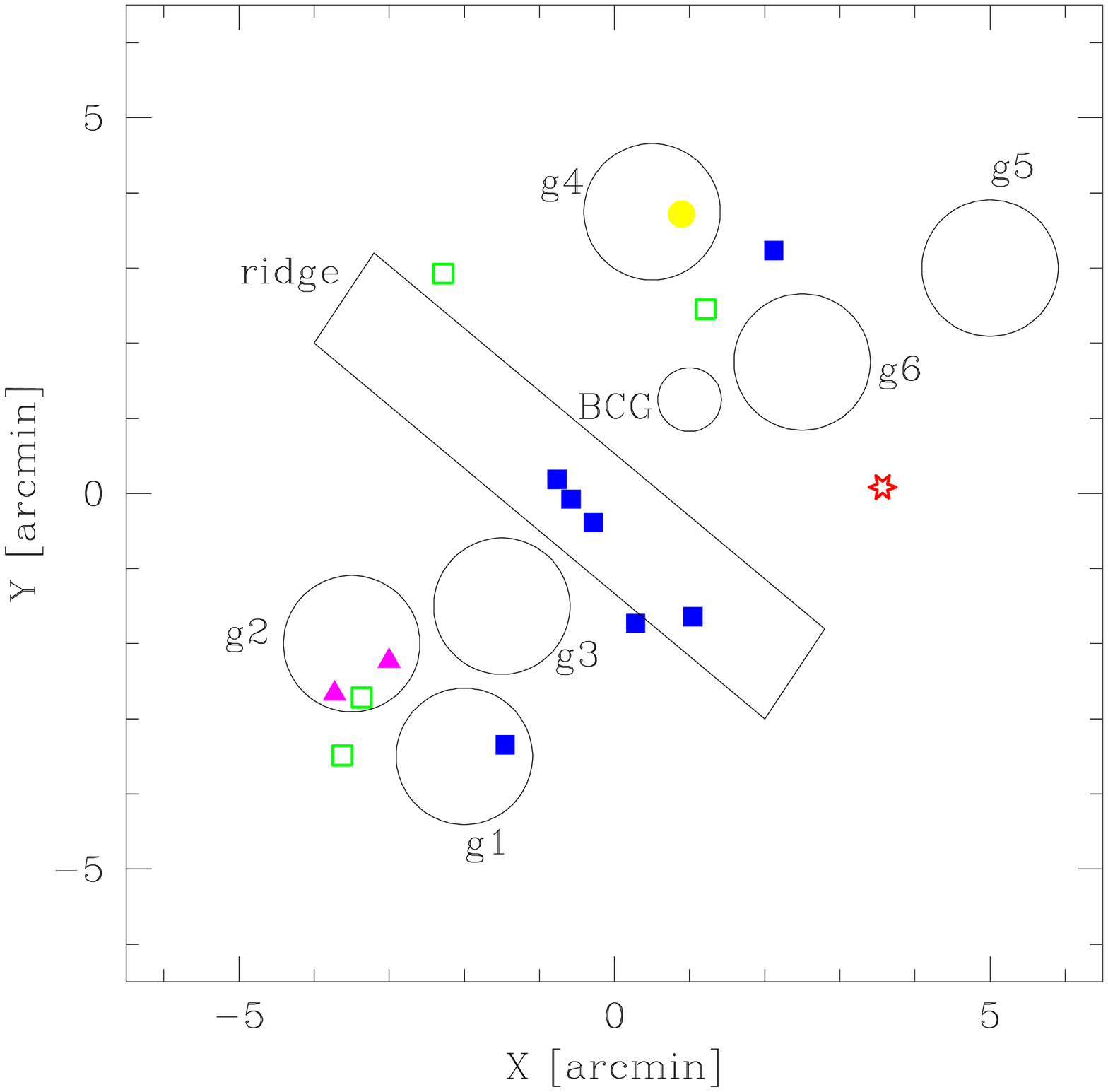}} 

\parbox[b]{16cm}{
\caption{Left: Velocity distribution (with a binning of 500 km/s) for galaxies with (white) and without (shading) emission lines. Right: Projected coordinates of the emission line galaxies.
{\bf Solid squares:} ${\rm v}_{r}=69000{\div}72000$~-~  
{\bf Open squares:} ${\rm v}_{r}=72000{\div}73000$~-~  
{\bf Solid circles:} ${\rm v}_{r}=74000{\div}75000$~-~
{\bf Stars:}  ${\rm v}_{r}=75000{\div}76000$~-~
{\bf Triangles:} ${\rm v}_{r}=76000{\div}80000$.
On the electronic version of the article, colors of the symbols have the same meaning as in Fig.~\ref{Slices}.}  
\label{noemi-emi}}  
\end{figure*}

\section{Discussion and Conclusions}

The velocity distribution of Abell 521 is definitively very complex. 
The large value of the velocity 
dispersion ($\sim$~1325 km/s) of our whole spectroscopical sample of 125
galaxies clearly results from the mixture of several components. The main
features of the velocity distribution are summarized hereafter:

i) a  high velocity tail in the velocity
distribution is revealed by a KMM partition. 
From the analysis of the 
projected positions and of the velocity/radius diagram, these objects are 
found to lie in the South-East region at about 870~\h~kpc from the X-ray
main center of the cluster, and to have velocities higher than 1500 km/s as
compared to the other objects at the same radius. The two-body
criteria shows that the probability that this
 system is not bound to the main
component of the cluster is quite high. 
These objects are probably field galaxies or a
loose background group. When excluding these galaxies, the mean
location remains unchanged ($\sim$~74000 km/s), but the scale reduces
slightly ($\sim$~1200 km/s).

ii) The velocity distribution in the high density central ridge shows 
a  velocity location  systematically lower than for the whole cluster, 
and a very high velocity dispersion
($1780 ^{+234} _{-142}$~km/s). Moreover the ratio of late/early type 
objects are higher than in the other slices 
(0.95 versus 0.46 in slice~1, and 0.48 in slice~3). 
This region is also particularly rich in emission-line galaxies, as it 
contains one third of the emission line objects detected in the whole
sample.
The late type, emission-line objects coincide 
with the low velocity tail of the 
distribution (v$<$72000~km/s, Figs.~\ref{plotN} and~\ref{noemi-emi}). 
Therefore, we are witnessing a very unusual 
configuration in the core of the cluster: 
a filamentary structure of ${\sim}~1~{{\rm h}_{75}}^{-1}$~Mpc, 
with galaxies showing a very broad velocity dispersion with an excess of
low velocity objects, colors typically bluer 
than the mean, and the presence of several emission line galaxies. 

iii) Various groups lie along the main NW/SE axis of the cluster. 
First, the region including the complex around the BCG ($\sim$~240~\h~kpc) 
appears as a strongly bound system 
with a very low velocity dispersion ($\sim~250$~km/s) typical of a
group and a location higher than the whole cluster($\sim~74340$~km/s). 
Other groups with comparable values of location are observed: g4 at
the NE extent and the southern group g3. 
The other groups are less well-sampled, so the following results have to
be taken with caution. The northern group g6 shows a slightly lower
location, while
at South, g1 seems to belong to a higher velocity complex, and g2
result of superposition effects. 

From these results, we can refine the scenario of formation of the
cluster. 
The  northern region is characterized by a lower velocity dispersion
and a slightly higher location; it hosts
a  group dynamically bound to the BCG; it is clearly   
associated to the compact group in X-ray, which is probably 
falling on the main cluster  (Arnaud~\etal 2000). 
The small difference in the mean velocity of the northern
region as compared to the whole cluster ($\sim~250$ km/s in
the rest-frame of the cluster) suggests that the merging occurs partly
in the plane of the sky, along the North-West/South East direction S2.
This direction  emerges as the main axis of the ongoing merging event.
Most  of the detected clumps are aligned along this direction. Moreover,
the early type objects as well as the brightest ones 
(L$>{\rm L}^*$) follow the general NW/SE skeleton of the cluster. 

However, our scenario has also to reproduce the peculiar features
that appear in the central region of the cluster, in particular its high
velocity dispersion, its lower location, and its filamentary NE/SW
structure. The high velocity dispersion is mostly due to the presence
of the low velocity
component detected by both the KMM partition and the velocity
profiles. It could consist of a foreground group 
currently interacting with the
main cluster. It is separated in radial velocity space 
by  $\sim~3000$~km/s from the main
cluster component. This high velocity bulk flow could result from a
recent merger with a  significant component along the line of sight
direction. Such high-speed encounters can be detected in merging clusters,
as shown for instance in the case of Cl0024+1654 (Czoske \etal 2002). 
The high density ridge S1 would then be the projection of the
merging axis along the plane of the sky. 
The large fraction of late type and star-forming objects 
along S1, coincident with the objects of the low velocity 
partition KMM-A, corroborates this hypothesis. In fact, a strong
compression of the gas   perpendicularly to the  
axis of the merger  between the two 
components is expected  during the merging event, which could trigger  
star formation in this region (Caldwell \etal 1993, Caldwell \& Rose 1997,   
Bekki 1999). Moreover, the brightest galaxy in the eastern side of the
ridge shows the same orientation as S1, and a velocity of $\sim$~72000~km/s. 
This object could in fact  be the original brightest galaxy of
the group which has collided the main cluster. 

These results imply that Abell 521 is the outcome of
multiple merger processes at various stages. A denser sampling in velocity
of the various groups, combined with a wider angular coverage are 
planned in order to understand the large-scale dynamics of this 
particular cluster.

\begin{acknowledgements} 
We warmly thank Monique Arnaud for intensive discussions, for her 
careful reading of the manuscript and
her comments which greatly improved the presentation of results.
We are very grateful to E.Slezak for performing one run of spectroscopy 
at ESO in 1999, leading to part of these data. We thank Antonaldo 
Diaferio and  Sandro Bardelli 
for fruitful discussion on the dynamics of clusters. 
Finally, we thank the referee, Alan Dressler, for his useful comments which 
helped us to improve and strengthen the paper. 
This work has been partially supported by the ``Programme National de 
Cosmologie'', by the Italian Space Agency grants
ASI-I-R-105-00 and ASI-I-R-037-01, and by the Italian Ministery (MIUR)
grant COFIN2001 ``Clusters and groups of galaxies: the interplay between
dark and baryonic matter".
\end{acknowledgements}

\begin{table*}[hbp]  
\begin{center}  
\caption{\small {\rm Velocity data}}  
\label {FindC}  
\begin{tabular}{llllllll} 
\hline  
\hline 
\multicolumn{8}{c}{ }\\ 
\# &
RUN &
R.A. &  
DEC. &  
HEL.VEL. &  
ERROR &  
Quality flag &  
Emission lines  \\  
 & & (2000) & (2000) & v(km/s) & $\Delta$v(km/s) & & \\ 
\multicolumn{8}{c}{ }\\ 
\hline 
\multicolumn{8}{c}{ }\\ 
1 & ESO2 & 04:53:47.91 & -10:11:23.52 & 89027 & 89 & 3 & NO \\ 
2 & ESO2 & 04:53:48.34 & -10:11:49.60 & 73101 & 39 & 1 & NO \\
3 & ESO1 & 04:53:49.70 & -10:13:32.42 & 33000 & 187 & 3 & NO \\
4 & ESO2 & 04:53:49.94 & -10:11:18.81 & 75683 & 51 & 1 & NO \\
5 & ESO1 & 04:53:50.26 & -10:12:19.00 & 25724 & 109 & 1 & H$\alpha$,H$\beta$,OIII \\
6 & ESO2 & 04:53:50.96 & -10:10:17.04 & 74317 & 100 & 3 & NO \\
7 & ESO1 & 04:53:51.55 & -10:12:42.90 & -2 & -2 & 1 & NO \\
8 & ESO2 & 04:53:52.01 & -10:11:52.27 & -2 & -2 & 1 & NO \\
9 & ESO1 & 04:53:52.05 & -10:12:31.30 & 74010 & 95 & 1 & NO \\
10 & ESO1 & 04:53:52.63 & -10:13:23.50 & -2 & -2 & 1 & NO \\
11 & ESO2 & 04:53:53.19 & -10:10:44.00 & -1 & -1 & 4 & NO \\
12 & ESO1 & 04:53:53.86 & -10:12:23.32 & 71668 & 124 & 2 & NO \\
13 & ESO2 & 04:53:54.46 & -10:12:29.41 & -2 & -2 & 1 & NO \\
14 & ESO1 & 04:53:54.75 & -10:12:19.65 & 74144 & 66 & 1 & NO \\
15 & ESO2 & 04:53:54.81 & -10:10:51.80 & -2 & -2 & 1 & NO \\
16 & ESO1 & 04:53:55.45 & -10:15:01.01 & 74017 & 73 & 1 & NO \\
17 & ESO2 & 04:53:56.40 & -10:10:02.15 & 74784 & 57 & 1 & NO \\
18 & ESO1 & 04:53:56.44 & -10:13:26.80 & 76181 & 139 & 1 & NO \\
19 & ESO1 & 04:53:56.56 & -10:14:46.29 & 75202 & 122 & 1 & NO \\
20 & ESO2 & 04:53:56.66 & -10:11:41.18 & -1 & -1 & 4 & NO \\
21 & ESO1 & 04:53:57.33 & -10:14:09.88 & 67749 & 86 & 2 & NO \\
22 & ESO1 & 04:53:57.82 & -10:14:24.45 & -1 & -1 & 4 & NO \\
23 & ESO2 & 04:53:58.10 & -10:12:09.25 & 74772 & 63 & 1 & NO \\
24 & ESO1 & 04:53:58.60 & -10:14:20.41 & 74584 & 80 & 1 & NO \\
25 & ESO1 & 04:53:58.77 & -10:13:25.85 & 71170 & 71 & 1 & NO \\
26 & ESO2 & 04:53:58.95 & -10:11:26.08 & 72694 & 38 & 1 & NO \\
27 & ESO2 & 04:53:59.29 & -10:13:01.14 & 121075 & 78 & 2 & NO \\
28 & ESO1 & 04:53:59.39 & -10:13:13.00 & -2 & -2 & 1 & NO \\
29 & ESO2 & 04:53:59.51 & -10:12:59.92 & 73455 & 34 & 1 & NO \\
30 & ESO1 & 04:53:59.93 & -10:13:01.63 & 73033 & 52 & 1 & NO \\
31 & ESO2 & 04:54:00.17 & -10:11:14.58 & 74250 & 40 & 1 & NO \\
32 & ESO1 & 04:54:00.27 & -10:14:05.13 & 74574 & 112 & 1 & NO \\
33 & ESO2 & 04:54:00.54 & -10:11:55.59 & 76219 & 49 & 1 & NO \\
34 & ESO1 & 04:54:00.70 & -10:13:09.70 & 182923 & 129 & 1 & OII \\
35 & ESO2 & 04:54:01.17 & -10:12:23.14 & 74657 & 114 & 1 & NO \\
36 & ESO2 & 04:54:01.41 & -10:11:16.30 & 71490 & 100 & 1 & OII,OIIIa,b,H$\beta$ \\
37 & ESO1 & 04:54:01.62 & -10:14:07.20 & 73928 & 67 & 1 & NO \\
38 & ESO1 & 04:54:01.75 & -10:12:34.47 & 99494 & 90 & 1 & OII,H$\beta$ \\
39 & ESO2 & 04:54:02.16 & -10:11:09.73 & 193062 & 100 & 3 & NO \\
40 & ESO2 & 04:54:02.26 & -10:12:49.86 & 99810 & 100 & 1 & OII,Balmer,H$\beta$ \\
41 & ESO1 & 04:54:02.34 & -10:14:04.40 & 70301 & 101 & 1 & NO \\
42 & ESO1 & 04:54:02.61 & -10:13:01.78 & 73195 & 92 & 2 & OII \\
43 & ESO2 & 04:54:02.97 & -10:11:29.06 & -2 & -2 & 1 & NO \\
44 & ESO1 & 04:54:03.26 & -10:13:22.40 & 39765 & 88 & 1 & NO \\
45 & ESO2 & 04:54:03.91 & -10:12:15.68 & 74900 & 111 & 1 & NO \\
46 & ESO1 & 04:54:04.02 & -10:14:42.44 & 73981 & 98 & 1 & NO \\
47 & ESO2 & 04:54:04.02 & -10:09:22.45 & -2 & -2 & 1 & NO \\
48 & ESO1 & 04:54:04.12 & -10:14:21.10 & -2 & -2 & 1 & NO \\
49 & ESO2 & 04:54:04.48 & -10:11:54.81 & -2 & -2 & 1 & NO \\
50 & ESO1 & 04:54:04.58 & -10:14:14.43 & 218349 & 70 & 2 & NO \\
\multicolumn{8}{c}{ }\\ 
\hline  
\end{tabular}  
\end{center}  
\end{table*}

\begin{table*}[hbp]  
\begin{center}  
\begin{tabular}{llllllll}  
\hline
\hline
\multicolumn{8}{c}{ }\\    
\# &
RUN &
R.A. &  
DEC. &  
HEL.VEL. &  
ERROR &  
Quality flag &  
Emission lines  \\  
 & & (2000) & (2000) & v(km/s) & $\Delta$v(km/s) & & \\
\multicolumn{8}{c}{ }\\   
\hline 
\multicolumn{8}{c}{ }\\ 
51 & ESO2 & 04:54:04.63 & -10:17:31.71 & 72713 & 62 & 1 & NO \\
52 & ESO1 & 04:54:04.90 & -10:12:03.68 & 72802 & 82 & 1 & OII,H$\beta$,OIIIa,b \\
53 & ESO1 & 04:54:05.06 & -10:16:04.27 & 69854 & 89 & 1 & H$\beta$,OIIIa,b \\
54 & ESO2 & 04:54:05.24 & -10:10:31.13 & 1 & 0 & 1 & QSO \\
55 & ESO1 & 04:54:05.48 & -10:14:10.80 & 74959 & 112 & 1 & NO \\
56 & ESO2 & 04:54:05.55 & -10:17:37.33 & 74862 & 71 & 1 & NO \\
57 & ESO1 & 04:54:05.60 & -10:11:15.50 & -2 & -2 & 1 & NO \\
58 & ESO2 & 04:54:05.79 & -10:11:41.71 & 74225 & 77 & 1 & NO \\
59 & ESO1 & 04:54:05.98 & -10:13:16.60 & 74571 & 50 & 1 & NO \\
60 & ESO2 & 04:54:06.20 & -10:12:39.05 & -2 & -2 & 1 & NO \\
61 & ESO2 & 04:54:06.36 & -10:10:49.94 & 74214 & 83 & 1 & OII,OIIIb \\
62 & ESO2 & 04:54:06.36 & -10:18:14.36 & 76171 & 59 & 1 & NO \\
63 & ESO1 & 04:54:06.37 & -10:15:59.98 & 109406 & 194 & 1 & OII,H$\beta$ \\
64 & ESO1 & 04:54:06.43 & -10:13:24.76 & 74340 & 100 & 2 & NO \\
65 & ESO1 & 04:54:06.56 & -10:13:21.32 & 74341 & 80 & 2 & NO \\
66 & ESO1 & 04:54:06.64 & -10:15:36.00 & 72394 & 66 & 2 & NO \\
67 & ESO2 & 04:54:06.65 & -10:11:06.49 & -1 & -1 & 4 & NO \\
68 & ESO1 & 04:54:06.94 & -10:11:46.43 & -2 & -2 & 1 & NO \\
69 & ESO1 & 04:54:07.02 & -10:16:04.20 & 44219 & 121 & 2 & NO \\
70 & ESO1 & 04:54:07.03 & -10:12:07.51 & 74475 & 79 & 1 & NO \\
71 & ESO1 & 04:54:07.06 & -10:17:55.90 & 74343 & 107 & 1 & NO \\
72 & ESO2 & 04:54:07.10 & -10:13:16.76 & 74205 & 106 & 1 & NO \\
73 & ESO1 & 04:54:07.12 & -10:12:41.25 & 110097 & 121 & 1 & OII,H$\beta$,OIIIa,b \\
74 & ESO1 & 04:54:07.14 & -10:15:11.36 & 74110 & 77 & 1 & NO \\
75 & ESO2 & 04:54:07.97 & -10:17:41.83 & -1 & -1 & 4 & NO \\
76 & ESO1 & 04:54:08.05 & -10:14:01.90 & 74334 & 70 & 1 & NO \\
77 & ESO2 & 04:54:08.11 & -10:10:20.47 & 74329 & 43 & 1 & NO \\
78 & ESO2 & 04:54:08.14 & -10:11:11.79 & 74536 & 79 & 1 & NO \\
79 & ESO1 & 04:54:08.23 & -10:14:58.59 & 68502 & 69 & 3 & NO \\
80 & ESO2 & 04:54:08.28 & -10:14:32.30 & 74369 & 100 & 2 & NO \\
81 & ESO1 & 04:54:08.31 & -10:17:13.59 & 55952 & 37 & 2 & H$\beta$ \\
82 & ESO1 & 04:54:08.41 & -10:12:41.95 & 75867 & 77 & 1 & NO \\
83 & ESO1 & 04:54:08.56 & -10:15:13.65 & 102858 & 116 & 3 & NO \\
84 & ESO1 & 04:54:08.60 & -10:15:50.10 & 75243 & 62 & 1 & NO \\
85 & ESO2 & 04:54:08.71 & -10:18:19.19 & 71848 & 42 & 1 & NO \\
86 & ESO2 & 04:54:08.83 & -10:10:48.42 & 74259 & 48 & 1 & NO \\
87 & ESO1 & 04:54:08.88 & -10:14:58.00 & 73236 & 64 & 1 & NO \\
88 & ESO2 & 04:54:08.99 & -10:11:06.96 & 109914 & 100 & 1 & OII \\
89 & ESO1 & 04:54:09.08 & -10:15:34.90 & 74438 & 81 & 1 & NO \\
90 & ESO1 & 04:54:09.33 & -10:15:10.08 & 40747 & 66 & 3 & NO \\
91 & ESO2 & 04:54:09.37 & -10:14:10.47 & 76738 & 85 & 1 & NO \\
92 & ESO1 & 04:54:09.47 & -10:17:11.80 & 75526 & 51 & 1 & NO \\
93 & ESO2 & 04:54:09.50 & -10:16:44.87 & 76202 & 114 & 2 & NO \\
94 & ESO1 & 04:54:09.86 & -10:16:03.13 & 72919 & 102 & 1 & NO \\
95 & ESO1 & 04:54:09.90 & -10:14:13.40 & 72972 & 47 & 1 & NO \\
96 & ESO1 & 04:54:09.90 & -10:17:34.20 & 78810 & 89 & 1 & NO \\
97 & ESO1 & 04:54:10.15 & -10:12:55.40 & -2 & -2 & 1 & NO \\
98 & ESO1 & 04:54:10.39 & -10:14:46.00 & 87413 & 90 & 1 & NO \\
99 & ESO1 & 04:54:10.58 & -10:15:10.54 & 106523 & 114 & 1 & OIIIa,b,H$\alpha$ \\
100 & ESO2 & 04:54:10.72 & -10:11:50.05 & -1 & -1 & 4 & NO \\
\multicolumn{8}{c}{ }\\ 
\hline  
\end{tabular}  
\end{center}  
\end{table*}

\begin{table*}[hbp]  
\begin{center}  
\begin{tabular}{llllllll}  
\hline
\hline
\multicolumn{8}{c}{ }\\   
\# &
RUN &
R.A. &  
DEC. &  
HEL.VEL. &  
ERROR &  
Quality flag &  
Emission lines  \\  
 & & (2000) & (2000) & v(km/s) & $\Delta$v(km/s) & & \\  
\multicolumn{8}{c}{ }\\ 
\hline
\multicolumn{8}{c}{ }\\  

101 & ESO2 & 04:54:11.06 & -10:17:35.16 & 75862 & 50 & 1 & NO \\
102 & ESO1 & 04:54:11.07 & -10:15:32.80 & -2 & -2 & 1 & NO \\
103 & ESO2 & 04:54:11.54 & -10:18:12.23 & -2 & -2 & 1 & NO \\
104 & ESO1 & 04:54:11.57 & -10:15:13.02 & 35443 & 50 & 2 & NO \\
105 & ESO2 & 04:54:11.72 & -10:14:35.28 & 70952 & 100 & 1 & OII,OIIIa,b,H$\beta$ \\
106 & ESO2 & 04:54:12.02 & -10:10:27.20 & -2 & -2 & 1 & NO \\
107 & ESO1 & 04:54:12.02 & -10:17:59.30 & -2 & -2 & 1 & NO \\
108 & ESO2 & 04:54:12.31 & -10:16:33.51 & -2 & -2 & 1 & NO \\
109 & ESO2 & 04:54:12.39 & -10:14:13.68 & 72076 & 60 & 1 & NO \\
110 & ESO2 & 04:54:12.50 & -10:14:20.31 & 70102 & 15 & 1 & OII,OIIIa,b,H$\beta$ \\
111 & ESO1 & 04:54:12.73 & -10:15:51.20 & 73965 & 36 & 1 & NO \\
112 & ESO1 & 04:54:13.02 & -10:15:51.80 & 72985 & 143 & 1 & NO \\
113 & ESO1 & 04:54:13.09 & -10:13:52.61 & 72067 & 100 & 1 & NO \\
114 & ESO2 & 04:54:13.10 & -10:14:17.53 & 36789 & 100 & 3 & NO \\
115 & ESO1 & 04:54:13.14 & -10:13:21.39 & 28080 & 59 & 1 & NO \\
116 & ESO1 & 04:54:13.19 & -10:16:08.94 & 93150 & 120 & 2 & NO \\
117 & ESO1 & 04:54:13.37 & -10:13:58.40 & 94388 & 113 & 2 & NO \\
118 & ESO1 & 04:54:13.39 & -10:15:17.74 & -1 & -1 & 4 & NO \\
119 & ESO2 & 04:54:13.44 & -10:11:17.66 & -2 & -2 & 1 & NO \\
120 & ESO2 & 04:54:13.54 & -10:17:27.27 & 131100 & 100 & 2 & OII \\
121 & ESO2 & 04:54:13.90 & -10:13:32.32 & 88714 & 100 & 1 & OII,Balmer \\
122 & ESO2 & 04:54:13.97 & -10:11:13.55 & 124290 & 100 & 3 & NO \\
123 & ESO1 & 04:54:13.97 & -10:18:02.64 & 76000 & 99 & 2 & NO \\
124 & ESO2 & 04:54:14.10 & -10:14:07.49 & 106911 & 89 & 2 & NO \\
125 & ESO1 & 04:54:14.21 & -10:13:43.40 & 107079 & 98 & 1 & NO \\
126 & ESO1 & 04:54:14.26 & -10:13:54.26 & 88284 & 94 & 2 & NO \\
127 & ESO2 & 04:54:14.53 & -10:14:43.21 & 89310 & 100 & 2 & OII,H$\beta$,OIIIb \\
128 & ESO1 & 04:54:14.55 & -10:12:35.33 & 73619 & 144 & 2 & NO \\
129 & ESO1 & 04:54:14.69 & -10:18:23.59 & 57498 & 70 & 3 & NO \\
130 & ESO2 & 04:54:14.77 & -10:17:47.29 & 70191 & 100 & 1 & OII,OIIIa,b,H$\beta$ \\
131 & ESO2 & 04:54:14.99 & -10:11:21.55 & 86628 & 54 & 1 & OII \\
132 & ESO1 & 04:54:15.03 & -10:13:48.87 & 127760 & 131 & 1 & NO \\
133 & ESO1 & 04:54:15.48 & -10:13:54.17 & 75229 & 90 & 1 & NO \\
134 & ESO2 & 04:54:15.62 & -10:17:20.94 & 130800 & 100 & 1 & OII,Balmer \\
135 & ESO2 & 04:54:15.68 & -10:10:09.49 & 109765 & 40 & 1 & NO \\
136 & ESO1 & 04:54:15.76 & -10:14:01.20 & 49153 & 124 & 2 & NO \\
137 & ESO1 & 04:54:15.81 & -10:16:47.36 & 71874 & 101 & 1 & NO \\
138 & ESO1 & 04:54:15.88 & -10:15:05.79 & 78411 & 55 & 2 & NO \\
139 & ESO1 & 04:54:16.01 & -10:16:11.60 & 74932 & 92 & 1 & NO \\
140 & ESO1 & 04:54:16.05 & -10:12:57.66 & 72704 & 93 & 1 & NO \\
141 & ESO2 & 04:54:16.08 & -10:10:08.86 & 73830 & 122 & 1 & NO \\
142 & ESO2 & 04:54:16.13 & -10:12:08.33 & -2 & -2 & 1 & NO \\
143 & ESO1 & 04:54:16.34 & -10:16:04.61 & 74282 & 61 & 1 & NO \\
144 & ESO1 & 04:54:16.44 & -10:15:09.60 & 73398 & 110 & 1 & NO \\
145 & ESO2 & 04:54:16.52 & -10:17:54.33 & 107203 & 84 & 2 & NO \\
146 & ESO1 & 04:54:16.60 & -10:14:30.60 & 73428 & 103 & 2 & NO \\
147 & ESO2 & 04:54:16.92 & -10:18:09.91 & 73070 & 31 & 1 & NO \\
148 & ESO1 & 04:54:17.12 & -10:14:25.17 & 71017 & 130 & 2 & NO \\
149 & ESO2 & 04:54:17.38 & -10:10:57.66 & 74409 & 76 & 1 & NO \\
150 & ESO2 & 04:54:17.54 & -10:13:05.01 & 98004 & 108 & 2 & NO \\
\multicolumn{8}{c}{ }\\
\hline  
\end{tabular}  
\end{center}  
\end{table*}

\begin{table*}[hbp]  
\begin{center}  
\begin{tabular}{llllllll}  
\hline
\hline
\multicolumn{8}{c}{ }\\  
\# &
RUN &
R.A. &  
DEC. &  
HEL.VEL. &  
ERROR &  
Quality flag &  
Emission lines  \\  
 & & (2000) & (2000) & v(km/s) & $\Delta$v(km/s) & & \\
\multicolumn{8}{c}{ }\\  
\hline
\multicolumn{8}{c}{ }\\

151 & ESO1 & 04:54:17.57 & -10:18:00.56 & 75611 & 88 & 1 & NO \\
152 & ESO1 & 04:54:17.93 & -10:18:04.53 & -2 & -2 & 1 & NO \\
153 & ESO2 & 04:54:18.17 & -10:10:30.22 & 74556 & 67 & 1 & NO \\
154 & ESO2 & 04:54:18.19 & -10:13:38.63 & 72587 & 75 & 1 & NO \\
155 & ESO1 & 04:54:18.39 & -10:13:22.62 & 31052 & 113 & 3 & NO \\
156 & ESO2 & 04:54:18.43 & -10:16:50.73 & 88515 & 91 & 1 & OII \\
157 & ESO2 & 04:54:18.56 & -10:13:39.25 & 97693 & 43 & 1 & OII,H$\beta$,OIIIa,b \\
158 & ESO1 & 04:54:18.87 & -10:18:12.17 & 74546 & 75 & 1 & NO \\
159 & ESO2 & 04:54:18.87 & -10:11:42.93 & 72600 & 121 & 1 & OII \\
160 & ESO2 & 04:54:18.92 & -10:15:16.89 & 74785 & 62 & 1 & NO \\
161 & ESO2 & 04:54:19.08 & -10:16:35.20 & 63328 & 96 & 2 & NO \\
162 & ESO1 & 04:54:19.23 & -10:15:19.14 & 92338 & 88 & 2 & NO \\
163 & ESO1 & 04:54:19.32 & -10:16:45.17 & 109151 & 74 & 1 & OII \\
164 & ESO1 & 04:54:19.75 & -10:13:49.60 & -2 & -2 & 1 & NO \\
165 & ESO2 & 04:54:19.78 & -10:11:27.08 & 75907 & 87 & 1 & NO \\
166 & ESO2 & 04:54:20.06 & -10:13:28.71 & 73571 & 91 & 1 & NO \\
167 & ESO1 & 04:54:20.53 & -10:10:22.20 & -2 & -2 & 1 & NO \\
168 & ESO1 & 04:54:20.65 & -10:17:26.06 & 99128 & 89 & 2 & NO \\
169 & ESO2 & 04:54:20.92 & -10:13:55.87 & -2 & -2 & 1 & NO \\
170 & ESO1 & 04:54:20.96 & -10:16:44.77 & 78326 & 146 & 1 & OII,H$\beta$ \\
171 & ESO2 & 04:54:21.12 & -10:09:39.91 & -2 & -2 & 1 & NO \\
172 & ESO1 & 04:54:21.29 & -10:12:07.41 & 101738 & 64 & 2 & NO \\
173 & ESO1 & 04:54:21.30 & -10:14:46.26 & 55236 & 86 & 2 & NO \\
174 & ESO2 & 04:54:21.63 & -10:12:02.65 & -1 & -1 & 4 & NO \\
175 & ESO1 & 04:54:21.65 & -10:16:41.02 & 73571 & 64 & 1 & NO \\
176 & ESO1 & 04:54:22.20 & -10:14:50.23 & -2 & -2 & 1 & NO \\
177 & ESO2 & 04:54:22.32 & -10:16:25.27 & 73694 & 51 & 1 & NO \\
178 & ESO1 & 04:54:22.35 & -10:17:13.40 & 72593 & 107 & 1 & H$\beta$ \\
179 & ESO1 & 04:54:22.64 & -10:16:40.67 & 79233 & 125 & 1 & NO \\
180 & ESO2 & 04:54:22.70 & -10:15:13.01 & 54106 & 71 & 1 & H$\beta$,OIIIa,b \\
181 & ESO2 & 04:54:23.18 & -10:17:58.06 & 72568 & 93 & 1 & OII,OIIIa,b,H$\beta$ \\
182 & ESO1 & 04:54:23.19 & -10:12:38.32 & -2 & -2 & 1 & NO \\
183 & ESO2 & 04:54:23.25 & -10:11:33.53 & 88370 & 59 & 1 & NO \\
184 & ESO1 & 04:54:23.73 & -10:13:05.92 & 109876 & 144 & 2 & NO \\
185 & ESO1 & 04:54:23.73 & -10:17:10.67 & 78122 & 100 & 1 & H$\beta$ \\
186 & ESO2 & 04:54:23.93 & -10:11:26.49 & 88524 & 50 & 1 & NO \\
187 & ESO1 & 04:54:24.16 & -10:14:00.74 & -2 & -2 & 3 & NO \\
188 & ESO1 & 04:54:24.55 & -10:16:40.88 & 73983 & 48 & 1 & NO \\
189 & ESO1 & 04:54:24.60 & -10:14:12.10 & 106221 & 88 & 1 & OII,H$\beta$,OIIIa,b \\
190 & ESO1 & 04:54:24.77 & -10:14:14.81 & 75310 & 84 & 3 & NO \\
191 & ESO1 & 04:54:24.91 & -10:16:58.01 & 72929 & 69 & 1 & NO \\   
\multicolumn{8}{c}{ }\\
\hline  
\end{tabular}  
\end{center}  
\end{table*}

\end{document}